\documentclass[a4paper,16pt]{article}

\usepackage{amsfonts}
\usepackage{graphicx,epstopdf}
\usepackage{caption}
\usepackage{subcaption}
\usepackage{siunitx}
\usepackage[colorlinks=true,linkcolor=blue,anchorcolor=blue,urlcolor=blue,citecolor=blue]{hyperref}
\DeclareGraphicsExtensions{.png,.eps,.pdf}
\usepackage{amsmath}
\usepackage{amsfonts}
\usepackage{latexsym}
\usepackage{amssymb}
\usepackage{verbatim}
\usepackage{multirow}
\usepackage{array}
\usepackage[percent]{overpic}
\usepackage{xcolor}
\usepackage{geometry}
\usepackage{framed}
\usepackage{todonotes}
\usepackage{hyperref}
\newcolumntype{L}{>{\centering\arraybackslash}m{5cm}}
\usepackage{colortbl}
\usepackage{booktabs}
\usepackage{eurosym}
\usepackage{pgfplotstable}

\usepackage{pgfplots}
\pgfplotsset{compat=newest}

\usepackage{tikz}
\usetikzlibrary{patterns}
\usetikzlibrary{shapes,arrows,calc}

\title{Experimental Analysis of Tip Vortex Cavitation Mitigation By Controlled Surface Roughness}

\usepackage{authblk}
\author[1]{U. Svennberg}
\author[2]{A. Asnaghi}
\author[1]{R. Gustafsson}
\author[2,a]{R.E. Bensow}
\affil[1]{Kongsberg Hydrodynamic Research Centre, Kongsberg Maritime AB, Kristinehamn, Sweden}
\affil[2]{Department of Mechanics and Maritime Sciences, Chalmers University of Technology}
\affil[a] {Author to whom correspondence should be addressed: rickard.bensow@chalmers.se}
\begin{document}
\maketitle
%\tableofcontents
 \section{Abstract}
This study presents results of experiments where roughness applications are evaluated in delaying the tip vortex cavitation inception of an elliptical foil. High-speed video recordings and Laser Doppler Velocimetry (LDV) measurements are employed to provide further details on the cavitation behaviour and tip vortex flow properties in different roughness pattern configurations. The angular momentum measurements of the vortex core region at one chord length downstream of the tip indicate that roughness leads to a lower angular momentum compared to the smooth foil condition while the vortex core radius remains similar in the smooth and roughened conditions. The observations show that the cavitation number for tip vortex cavitation inception is reduced by 33 \% in the optimized roughness pattern compared to the smooth foil condition where the drag force increase is observed to be around 2 \%. During the tests, no obvious differences in the cavitation inception properties of  uniform and non-uniform roughness distributions are observed. However, the drag force is found to be higher with a non-uniform roughness distribution.

\noindent
\textbf{Keywords:} Roughness, Tip vortex, Mitigation, Cavitation, Suppression

 \section{Introduction}
The physics of tip vortex cavitation, TVC, involves simultaneous presence of small flow structures and phase change. Since controlling TVC is vital for low-noise propellers \cite{Pennings20151,Mukund2016,Zhang2015488}, intensive efforts have been conducted both experimentally and numerically to understand the physics of this flow. The marine industry generally seeks only to increase the tip vortex core pressure to alleviate cavitation, and to do so, a few approaches are proposed and tested. These approaches generally can be classified into two groups. In the first group, which is called active control, the tip vortex flow is changed by injection of a solution, e.g. air, polymer, or water. In the second group so called passive control, either the propeller geometry is modified or its surface properties, e.g. roughness. 

\noindent   %Injection:Done
In the injection method, interaction between the injected solution and the original tip vortex defines the mitigation. Consequently, the location of injection, the type and amount of solution are decisive parameters ~\cite{Chahine1993497,Rivetti2016}. As an example, Chahine et al. ~\cite{Chahine1993497} observed up to 35 \% delay in TVC inception (TVCI) by injecting some combinations of polymer solutions while there was not any significant improvement in TVCI delay by the injection of pure water or a 50 \% water-glycerin viscous mixture. 

\noindent   %Adding extra geometry:Done
Another approach to weaken a tip vortex is to include an extra geometry, e.g. endplate or winglet, attached to the tip. The idea is to stabilize the Reynolds shear stress and therefore, reduce the tip vortex strength ~\cite{Gim201328,Amini2019}. For tip loaded propellers (TLP), having winglet prevents formation of a single strong tip vortex, and instead two or in some cases a few distinct weaker vortices are generated ~\cite{Brown2015}. It is expected that the evolution of the tip vortices and their interaction defining the mitigation depend on a number of factors, some of those being the span of the endplate and the loading distribution on it. 

\noindent 
Park et al. ~\cite{Park20141} proposed a semi-active control scheme tested on a model scale propeller by attaching a flexible thread at the tip of the blade. Due to a low pressure region of the tip vortex core, the thread is sucked into the tip vortex. Three different thread materials were tested: steel, nylon, and Dyneema, where only the Dyneema thread was sufficiently flexible to be sucked into the tip vortex core. It is reported that when the thread is sucked into the vortex, it can effectively suppress the tip vortex cavitation inception. Amini et al. ~\cite{Amini2019b} investigated the effectiveness of a flexible nylon thread in TVC mitigation of an elliptical hydrofoil. They investigated several thread sizes (thickness and length) and conclude that certain configurations could remarkably attenuate TVC without imposing any tangible penalties on the hydrodynamic performances. Their conclusion highlights that the thread sizes should be selected according to the flow conditions in a way the thread becomes unstable and interacts with the vortex dynamically to mitigate TVC. The study of Lee et al. ~\cite{Lee2018} suggests the presence of the thread at the tip causes a significant reduction in the streamwise velocity field around the vortex core region and therefore mitigates TVC. 

\noindent   %Modifying geometry:Done 
The tip geometry can be modified as an effort to restrain sheet and tip vortex cavitation in the design of marine propellers ~\cite{ASNAGHI2018197, AsnaghiPhDThesis}. This, however, can cause cloud cavitation which can give noise and surface erosion on tip-modified propellers ~\cite{Shin2015}.

\noindent  %Roughness:Done 
Roughening the blade, which is the main topic of this study, will also lead to tip vortex radius increase, and therefore delay in TVCI ~\cite{McCormick1962369}. This is believed to be related to the turbulence generated by the roughness which destabilizes the tip vortex and leads to its early breakdown. As discussed by Souders and Platzer \cite{bhl102158}, a thicker turbulent boundary layer on the wing tip resulted from the roughness increases tip vortex core radius and decreases the tip vortex maximum tangential velocity as compared to the laminar flow case. The surface roughness also contributes to other tip vortex properties, e.g. the roll-up process \cite{Katz89}. In the study of Johnsson and Ruttgerson \cite{Johnsson91}, the influence of leading edge roughness on the tip vortex roll-up for different angles of attack was investigated. They concluded that application of roughness on the pressure side near the leading edge has a delaying effect on tip vortex cavitation and leads to a substantial reduction in tip vortex strength. They also noted an open water efficiency degradation when roughness was applied, around 2$\%$ in their tested condition. Later, it was suggested by Philipp and Ninnemann \cite{Philipp2007} to apply the roughness only on a small area of the blade suction side close to the tip trailing edge region, i.e. in the area of the tip vortex formation. This was tested by Kruger et al. \cite{Kruger2016110} on a propeller where different roughness patterns were applied on different parts of a propeller to find the most optimum location for roughness.

\noindent 
Even though that experimental tests have proven the capability of using roughness to mitigate tip vortex flows and delay cavitation, this method has not become practically available yet. The main reason is the lack of knowledge on how exactly roughness mitigates tip vortex. The required or optimum roughness properties such as height, distribution, type, and pattern are the main decisive parameters. Moreover, it should be clarified whether it is increasing the boundary layer thickness or providing further turbulent instabilities that leads to tip vortex mitigation. Last but not least, the scale law on how model scale roughness studies can be extended to the full scale propellers should be developed \cite{Park2017962}. The current study has been performed within a project aimed to find the answers of these questions. To add to the knowledge, experiments for tip vortex flows and cavitation inception around an elliptical foil have been conducted. 

\noindent 
The tip vortex flow and cavitation inception around smooth elliptical foils have been investigated by several researchers ~\cite{Arndt1995,Arndt2002143,PenningsThesis,AsnaghiSMP2015,Peng2017939}. This type of foils is of interest as the vortex structures around it resembles the propeller tip vortex behaviour while making it possible to be tested in more details both experimentally and numerically. While the previous studies were focused on the formation and development of tip vortex over smooth surfaces, the current measurements include the roughness application and how it changes the tip vortex properties. In order to provide more details for analysis, the geometry of the elliptical foil tested at Delft Technical University is used ~\cite{Pennings20151,Pennings2015288}. However, due to the velocity speed limitation of our cavitation tunnel, the current foil has a scale ratio of 2.398 compared to the one tested at Delft TU. Our recent numerical analysis ~\cite{asnaghiSMP2019} is used as an initial guideline for the pattern and size of the roughness elements. The main objective here is to evaluate these patterns in order to provide more insights on the roughness impact in tip vortex inception and cavitating tip vortex behaviour.

\noindent 
The tests are conducted within a wide range of tip vortex conditions from non-cavitating to fully cavitating tip vortex along with varying the free-stream velocity that is considered to evaluate the impact of Reynolds number. In order to include the effects of the sand grains irregularities, in one test the roughness pattern is repeated with different distributions of sand grains on the foil. The outcome of these tests provides further knowledge on how roughness should be applied to achieve the highest TVC mitigation with a minimized negative impact on hydrodynamic performance. Cavitation inception was recorded through visual inspection of high-speed recordings.

\section{Experimental setup}
 \noindent
The experiments are performed in the free surface cavitation tunnel at the Kongsberg Hydrodynamic Research Center, Kristinehamn, Sweden. The cross section of the test section is $0.8 \times 0.8$ $\text{m}^2$ at the inlet and $0.8 \times 0.82$ $\text{m}^2$ at the outlet. The vertical direction was extended gradually from inlet to outlet to compensate for the growth of the boundary layer in order to facilitate a nearly zero-pressure gradient in the streamwise direction. This is provided by having the top plate horizontal and the bottom plate with a downward slope of 0.4 degrees where at the foil section, the vertical distance is about $0.81$ m. 

\noindent
The tested foil is a half-model wing of elliptic planform with a NACA $66_2 -415$ cross section. Compared to the elliptical foil recently tested at Delft TU ~\cite{PenningsThesis,Pennings2015288}, the current foil has a scale ratio of 2.398 leading to the root chord length $\text{C}_0=301.2$ mm and the span length of S=$360$ mm. The projected surface area of the foil computed from the CAD file is A=$8.43 \times 10^{-2}$ $\text{m}^2$. The lift and drag coefficients are calculated based on this area, $\text{C}_\text{l}=\text{L}/(0.5 \rho {\text{U}_\text{inlet}}^2 A)$ and $\text{C}_\text{d}=\text{D}/(0.5 \rho {\text{U}_\text{inlet}}^2 A)$ where L and D are the lift and drag forces.

\begin{figure} [h!]
        \centering
		\includegraphics[width=0.6\textwidth]{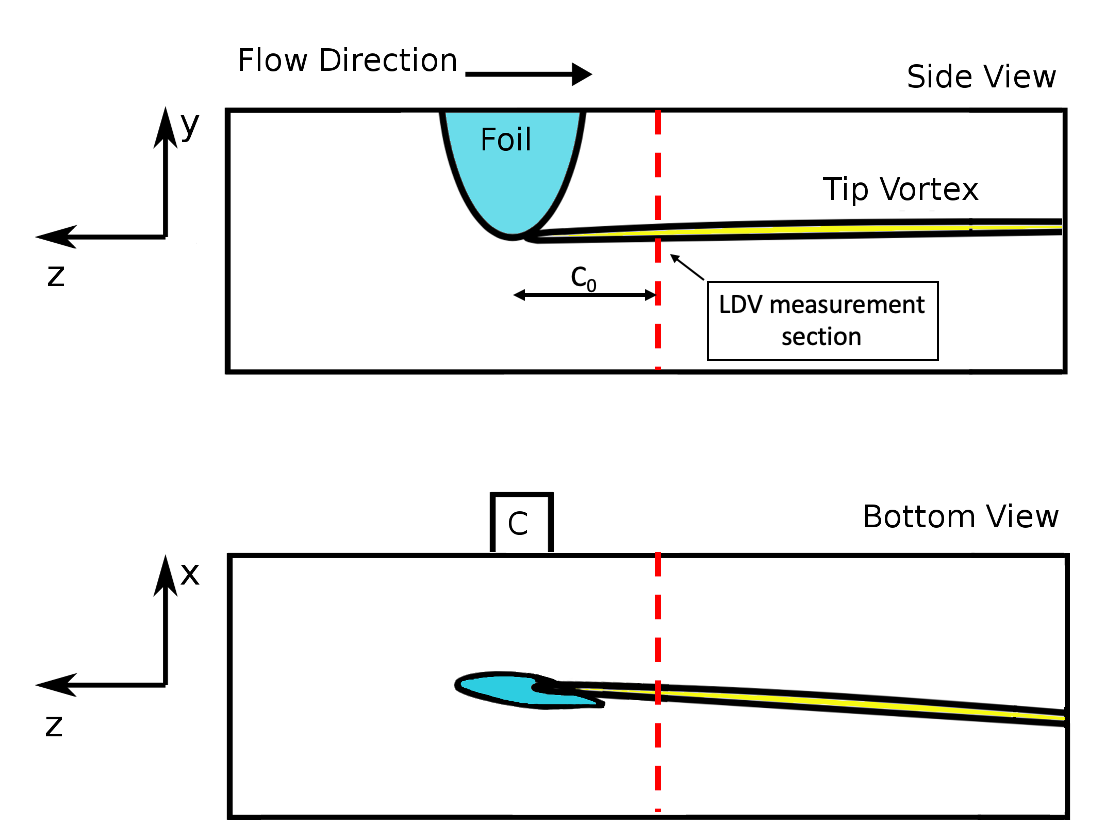}   
		\caption{Sketch of the test tunnel and the coordinate system.}
		\label{fig::coordinateSystem}
\end{figure}

\noindent
The sketch of the test setup is given in Fig. \ref{fig::coordinateSystem}. The coordinate system, with the foil tip at the origin, is defined with z pointing towards the inlet. The foil is mounted on a rotatable shaft through the top plate having 1 mm gap between the root section of the foil and the top plate to give the possibility of varying the angle of attack. A five-component force/torque sensor (Kongsberg dynamometer no. 54A) is mounted between the shaft and the foil to measure the forces. The location of the camera is also highlighted in the bottom view. The lift and drag forces are reported in the y and z directions, respectively. The high-speed camera (FASTCAM SA-X2) is placed near the side window while viewing the suction side of the foil. 

\noindent
The measurements are performed with respect to the averaged free-stream velocity ranging from 2.84 m/s to 4 m/s with typical fluctuations of $\pm 0.7 \%$. The outlet cavitation number is varied from 2.6 to 8 while keeping the inflow velocity constant. The Reynolds number is calculated based on the root chord length and the inlet flow properties, Re=$\rho \text{U}_{\text{inlet}} \text{C}_0 / \mu$, and the cavitation number is calculated based on the outlet pressure, $\sigma=(\text{p}_{\text{outlet}}-\text{p}_{\text{sat}})/(0.5 \rho \text{U}_{\text{inlet}})$. The operating conditions were monitored during each experiment to prevent any deviations during the measurements. The water temperature, kept around 23 $^{\circ}$C during the tests, is measured with a PT-100 sensor placed submerged in the tunnel water downstream of the test section.  

\begin{figure} [h!]
        \centering
		\includegraphics[width=0.6\textwidth]{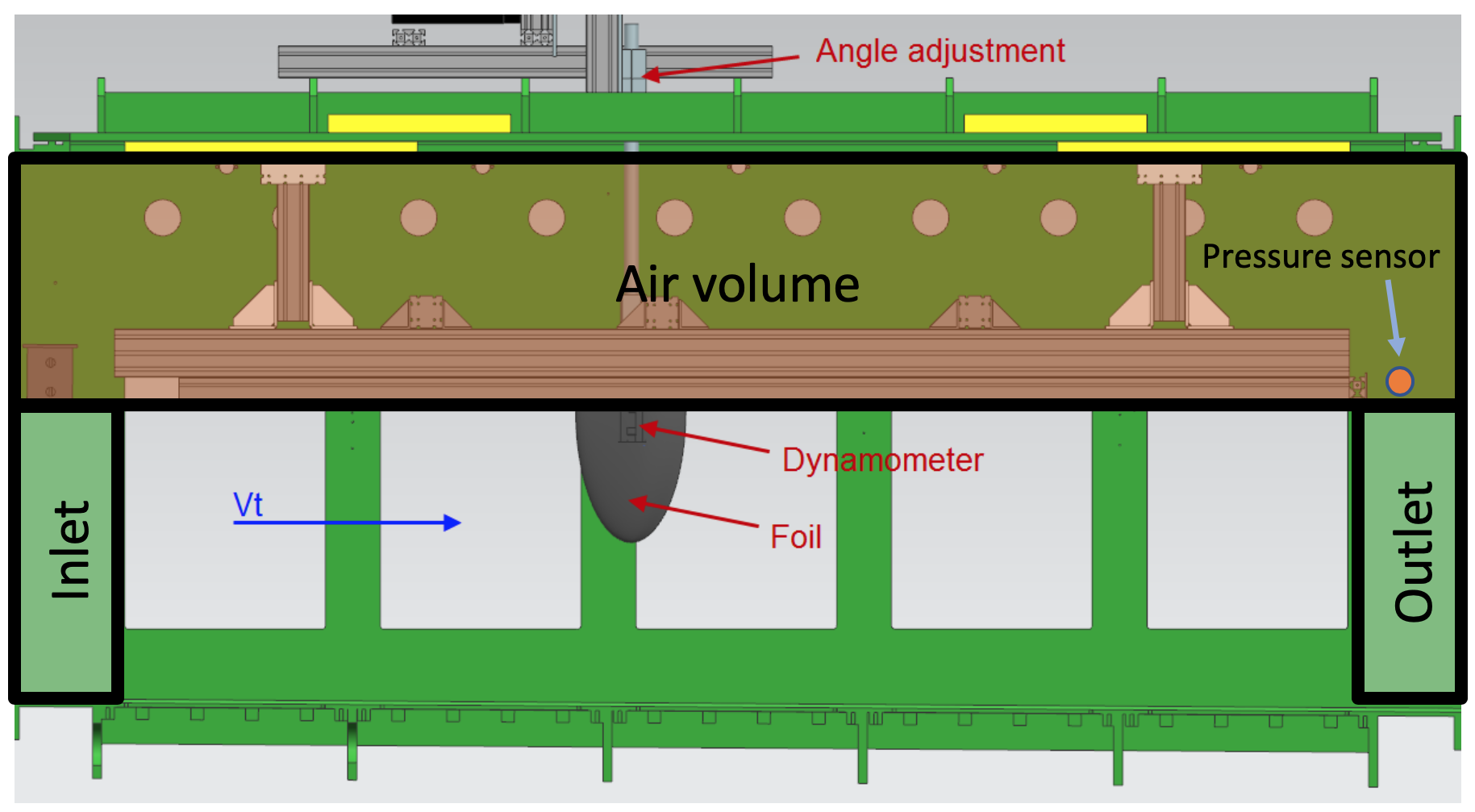}   
		\caption{Cavitation tunnel sketch and the location of the measurement equipments.}
		\label{fig::coordinateSystemCAD}
\end{figure}

\noindent
Static pressure was measured with a low pass filter at about $0.2$ Hz with a digital absolute pressure sensor (Rosemount 3051S). The pressure was measured in the air volume and compensated for the water depth down to the tip of the foil, Fig. \ref{fig::coordinateSystemCAD}. The typical accuracy is $0.05 \%$ of full-scale pressure ($1.1 \times 10^5$ Pa), which is 55 Pa. The free-stream velocity is measured using a pitot tube having the same type of the pressure transducers and located upstream of the wing. 

\noindent
The dissolved oxygen concentration (DO) was used as a measure of the amount of dissolved gas in the water using a fluorescence-based optical sensor (Trioxmatic 690). It has a range of 0-600 \% which equals to 0-60 mg/l and the measurements were conducted at around 40 \% which corresponds to 4 mg/l. As only dissolved oxygen is measured, this is taken as a representative indicator for the total amount of dissolved gas.

\section{Tested conditions}
\begin{figure} [h!]
        \centering
		\includegraphics[width=0.58\textwidth]{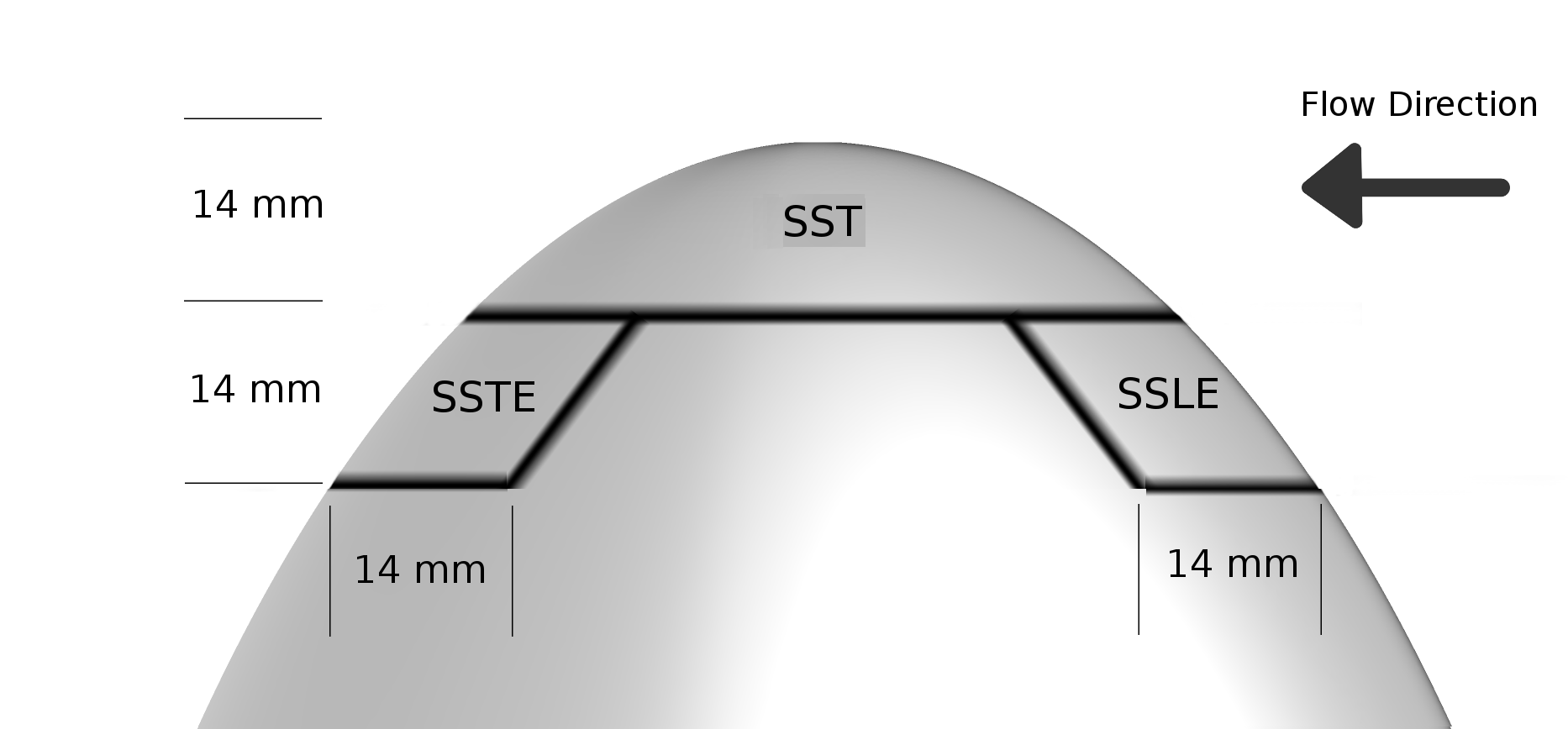}   
		\caption{Description of the areas where the roughness is applied, zoomed view, suction side of the foil.}
		\label{fig::testedConditionsRoughnessPattern}
\end{figure}

\begin{figure} [h!]
        \centering
        \begin{subfigure}[b]{0.325\textwidth}
        		  \begin{tikzpicture}
				  \node [anchor=south west,inner sep=0](mesh)at(0,0){
				  \includegraphics[width=\textwidth]{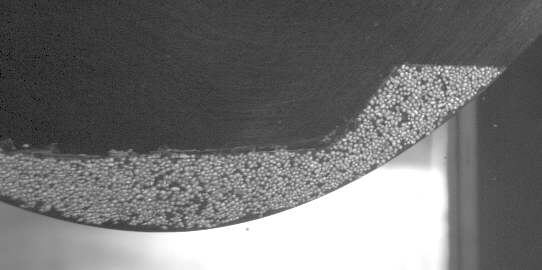} 
									};
									
					\node [anchor=south west,inner sep=0, black] (mesh) at (6.1,3.5) {\Large Free-stream direction};	
					\draw [red, very thick, ->]  (5.7,3.25)  ->  (11.4,3.25);  
				  \end{tikzpicture}	 
				  				  \caption{Uniform distribution}
                  \label{fig::uniformDisSandGrains}
        \end{subfigure}           
        \begin{subfigure}[b]{0.325\textwidth}
				  \includegraphics[width=\textwidth]{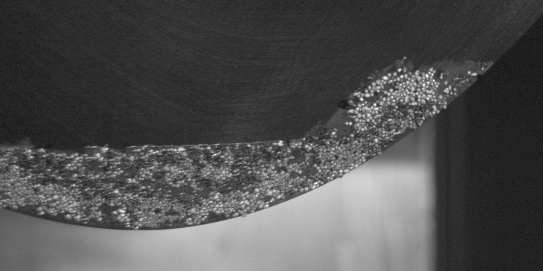}; 
				  \caption{Non-uniform distribution}				  
                  \label{fig::nonUniformDisSandGrains}
        \end{subfigure} 
        \begin{subfigure}[b]{0.325\textwidth}
				  \includegraphics[width=\textwidth]{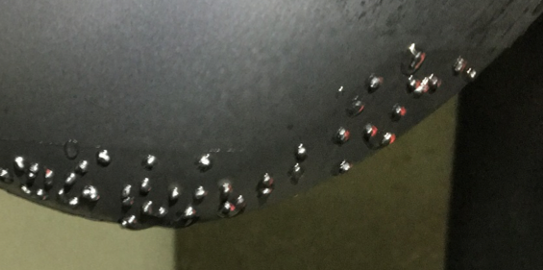}; 
				  \caption{Sparse distribution}				  
                  \label{fig::nonUniformDisSandGrains}
        \end{subfigure}                
		\caption{Different distributions of sand grains for the roughness pattern (II).}
		\label{fig::uniformAndNonUniformDisSandGrains}
\end{figure}

 \noindent
Previous LES analysis of the tip vortex flow around the foil indicates that for considered conditions three different areas on the suction side are important in the tip vortex formation and development ~\cite{asnaghiSMP2019}. These findings are concluded based on the evaluation of the tip vortex flow streamlines and momentum of vortical structures fed into the tip vortex core. It has been discussed that the side of the foil where the roughness should be applied is the side that the vortex roll-up forms, e.g. the suction side for the current operating condition. In our later study ~\cite{asnaghiPOF2020}, the impact of the roughness area on the flow streamlines and separation lines are investigated. It is highlighted that roughness provides an extra separation line close to the tip along with the leading edge separation line on the side where the roll-up forms. Due to the resource limitation, the roughness experimental tests are focused only on these areas where the height is set equal to the optimum height found in the numerical analysis, equal to 230 $\mu$m. The first roughness pattern, called pattern (I) in this paper, consists of an area on the leading edge (SSLE) and an area on the tip of the foil (SST). In the roughness pattern (II), an extra area on the trailing edge (SSTE) is also included. The description and sizes of these areas are presented in Fig. \ref{fig::testedConditionsRoughnessPattern}. In order to include the effects of the sand grains irregularities, the tests on the pattern (II) are repeated with different distributions of sand grains on the foil, Fig. \ref{fig::uniformAndNonUniformDisSandGrains}. 

\noindent
The cavitating tip vortex is evaluated at the cavitation numbers 1.2 and 2.6. The tip vortex cavitation inception and development are evaluated at cavitation numbers larger than 3.8. Most of the tests are carried out at the free-stream velocity of 2.84 m/s. A few of the conditions is repeated at a higher inlet velocity, i.e. 4 m/s, to evaluate the Reynolds number impact on the tip vortex development. It also provides the opportunity of evaluating the roughness impact on the risk for sheet cavitation at different velocities. 
\section{Results and discussion}
 \subsection{Force measurements}
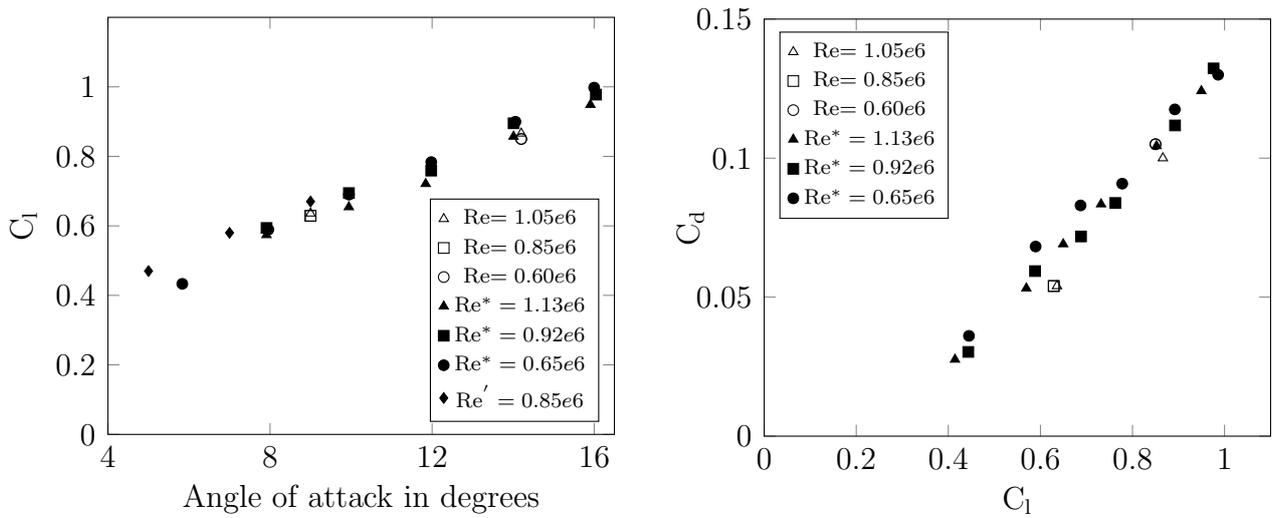
\begin{figure}[h!]
    \centering
		\begin{subfigure}[b]{0.485\textwidth}
		  \centering
		  \selectcolormodel{gray}
			\begin{tikzpicture}
			  \begin{axis}[font=\large,legend pos=south east, xlabel=Angle of attack in degrees, ylabel=$\text{C}_\text{l}$, width=\textwidth, xmin=4, xmax=16.5, ymin=0, ymax=1.2, xtick={0,4,8,12,16}, ytick={0,0.2,0.4,0.6,0.8,1.0}]
			  
			  \pgfplotsset{every x tick label/.append style={font=\large, yshift=-0.25ex}}
			  \pgfplotsset{every y tick label/.append style={font=\large, xshift=-0.25ex}}

			  	\addplot[only marks,mark=triangle,mark size=2] 	table[only marks, x index=0,y index=1]  {PLOT/SmoothFoil/ClApha.dat};
				\addplot[only marks,mark=square,mark size=2] 	table[only marks, x index=2,y index=3]  {PLOT/SmoothFoil/ClApha.dat};
				\addplot[only marks,mark=o,mark size=2] 		table[only marks, x index=4,y index=5]  {PLOT/SmoothFoil/ClApha.dat};

			  	\addplot[only marks,mark=triangle*,mark size=2] table[only marks, x index=0,y index=1]  {PLOT/SmoothFoil/ClAlphaExp113.dat};	
			  	\addplot[only marks,mark=square*,mark size=2] 	table[only marks, x index=0,y index=1]  {PLOT/SmoothFoil/ClAlphaExp92.dat};			  				  	\addplot[only marks,mark=otimes*] 					table[only marks, x index=0,y index=1]  {PLOT/SmoothFoil/ClAlphaExp65.dat};

				\addplot[only marks,mark=diamond*,mark size=2] 	table[only marks, x index=0,y index=1]  {PLOT/SmoothFoil/ClAphaPennings.dat};		
							  					
				\addlegendentry{\footnotesize Re$=1.05e6$}
				\addlegendentry{\footnotesize Re$=0.85e6$}        
				\addlegendentry{\footnotesize Re$=0.60e6$}  

				\addlegendentry{\footnotesize Re$^*=1.13e6$}        				      
				\addlegendentry{\footnotesize Re$^*=0.92e6$}        				      
				\addlegendentry{\footnotesize Re$^*=0.65e6$}  
				\addlegendentry{\footnotesize Re$^{'}=0.85e6$}        				      				      				      
																								
			  \end{axis}
			\end{tikzpicture}
		\end{subfigure}		
		\hspace{0.2cm} 		
		\begin{subfigure}[b]{0.485\textwidth}
		  \centering
		  \selectcolormodel{gray}
			\begin{tikzpicture}
			  \begin{axis}[font=\large,legend pos=north west, xlabel=$\text{C}_\text{l}$, ylabel=$\text{C}_\text{d}$, width=\textwidth, xmin=0, xmax=1.1, ymin=0, ymax=0.15, xtick={0,0.2,0.4,0.6,0.8,1.0}, ytick={0,0.05,0.1,0.15}, yticklabel style={/pgf/number format/fixed,/pgf/number format/precision=2},y label style={at={(-0.1,0.5)}}]
			  
			  \pgfplotsset{every x tick label/.append style={font=\large, yshift=-0.25ex}}
			  \pgfplotsset{every y tick label/.append style={font=\large, xshift=-0.25ex}}

			  	\addplot[only marks,mark=triangle,mark size=2] 					table[only marks, x index=0,y index=1]  {PLOT/SmoothFoil/CdCl.dat};
				\addplot[only marks,mark=square,mark size=2] 	table[only marks, x index=2,y index=3]  {PLOT/SmoothFoil/CdCl.dat};
				\addplot[only marks,mark=o,mark size=2] 		table[only marks, x index=4,y index=5]  {PLOT/SmoothFoil/CdCl.dat};				

			  	\addplot[only marks,mark=triangle*,mark size=2] table[only marks, x index=0,y index=1]  {PLOT/SmoothFoil/CdClExp113.dat};	
			  	\addplot[only marks,mark=square*,mark size=2] 	table[only marks, x index=0,y index=1]  {PLOT/SmoothFoil/CdClExp92.dat};			  				  	\addplot[only marks,mark=otimes*] 					table[only marks, x index=0,y index=1]  {PLOT/SmoothFoil/CdClExp65.dat};
			  					
				\addlegendentry{\footnotesize Re$=1.05e6$}
				\addlegendentry{\footnotesize Re$=0.85e6$}        
				\addlegendentry{\footnotesize Re$=0.60e6$}  

				\addlegendentry{\footnotesize Re$^*=1.13e6$}        				      
				\addlegendentry{\footnotesize Re$^*=0.92e6$}        				      
				\addlegendentry{\footnotesize Re$^*=0.65e6$}        				      
																								
			  \end{axis}
			\end{tikzpicture}
		\end{subfigure}	 
    \caption{Variation of the lift and drag coefficients for different angles of attack and Reynolds numbers, smooth foil, non-cavitating condition, Re: current tests, Re$^*$: extracted from Arndt and Keller ~\cite{Arndt1992}, and Re$^{'}$: extracted from Pennings ~\cite{PenningsThesis}.}
    \label{fig:CdClAlphaVariationSmooth}
\end{figure}

 \noindent
The first part of the results contains the smooth foil measurements at different Reynolds numbers and angles of attack, Fig. \ref{fig:CdClAlphaVariationSmooth}. The results are elaborated by providing comparisons with the measurements conducted by Arndt and Keller ~\cite{Arndt1992} and Pennings ~\cite{PenningsThesis}. It should be noted that these tests are conducted on the similar foil having slightly different geometrical aspect ratio where the root chord length of the foil used in ~\cite{Arndt1992} was equal to $\text{C}_0=129.4$ mm, and in ~\cite{PenningsThesis} was equal to $\text{C}_0=125.4$ mm. Lift data are collected over a range of Reynolds numbers from $0.6 \times 10^6$ to $1.13 \times 10^6$, and the angles of attack ranging from 5$^\circ$ to 16$^\circ$. Each measurement point is averaged over at least five measured samples having around 30 s duration. The measurements variation of each point is found to be less that 2\% which to avoid the ambiguity is not included in the figure.

\noindent
In general, the measured lift and drag coefficients follow expected trends. The differences can be related to the uncertainties of the measurements and the devices used in different cavitation tunnels.

\subsection{Flow field}
\begin{figure} [h!]	
        \centering
        \begin{subfigure}[b]{0.45\textwidth}
				  \includegraphics[width=\textwidth]{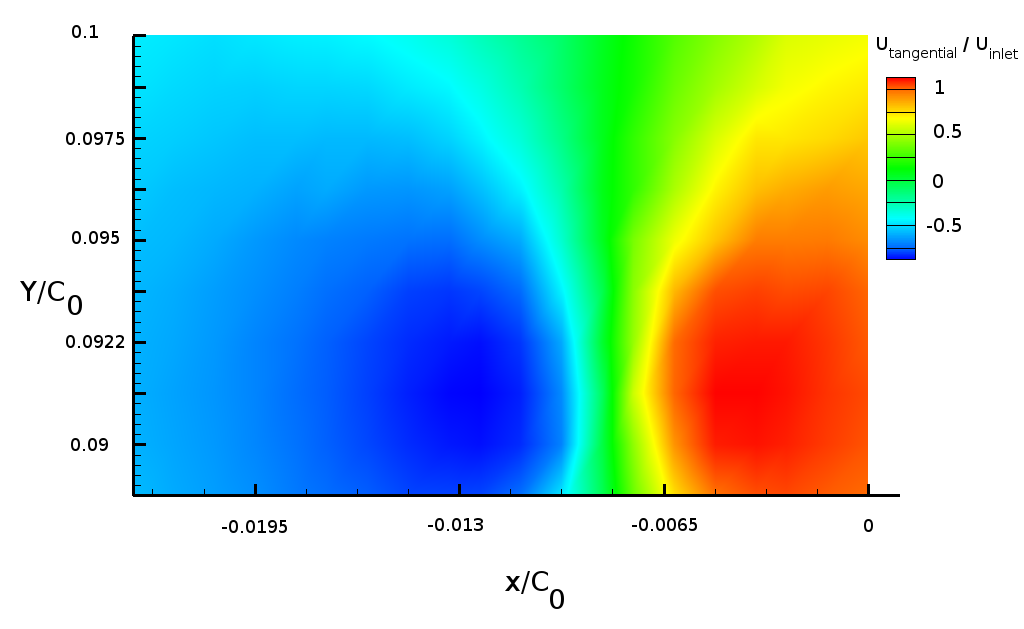}
				  \caption{Vertical velocity}
                  \label{fig::SmoothVerticalPlot}
        \end{subfigure}
        \begin{subfigure}[b]{0.45\textwidth}
				  \includegraphics[width=\textwidth]{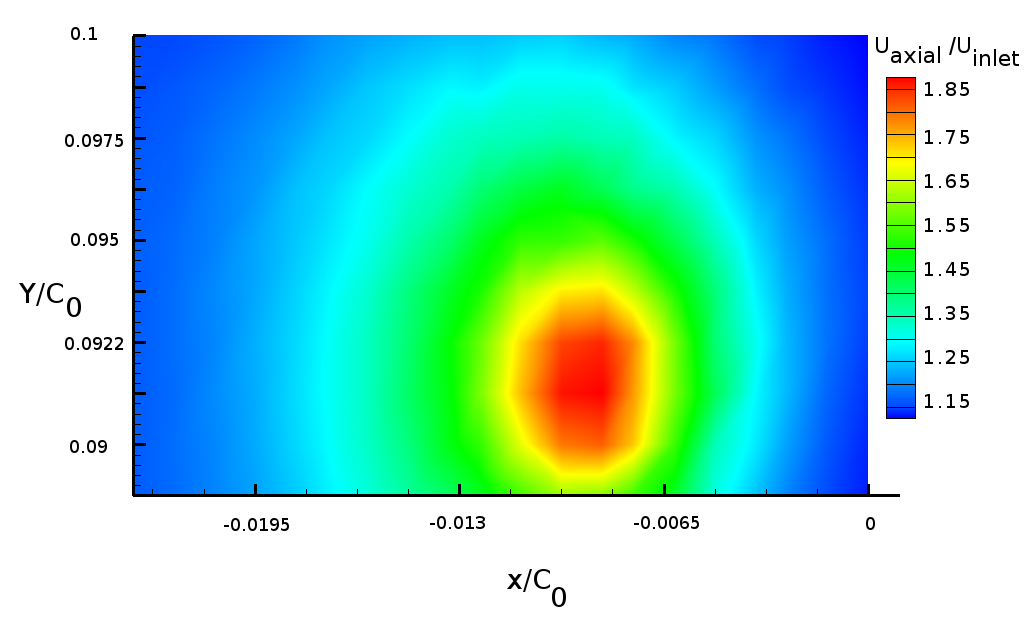}
				  \caption{Axial velocity}
                  \label{fig::SmoothAxialPlot}
        \end{subfigure}  
		\caption{Axial and vertical velocity (velocity in y-direction) distributions for smooth foil at non-cavitating condition, Re=$1.204 \times 10^6$, $\alpha$= 9$^\circ$, z/$\text{C}_0$=1.}          
		 \label{fig::SmoothVelPlot}                   	
\end{figure}

 \noindent
The velocity distribution at one chord length downstream of the smooth foil measured by LDV is presented in Fig. \ref{fig::SmoothVelPlot}. The measurements are conducted at the inlet velocity of 4 m/s, equal to Re=$1.204 \times 10^6$. To provide around 20 measurement points across the vortex core radius, the measurement resolution of 60 $\mu$m is applied over the vortex core region. The measurement resolution, then, gradually becomes coarser outside the vortex core where the velocity gradients are smaller.

\noindent
Close to the foil, the tangential velocity is highly asymmetric but gradually becomes symmetric further downstream as the roll-up process develops. This asymmetric behaviour can be noted from the velocity distribution in the y-direction (noted as vertical velocity), Fig. \ref{fig::SmoothVerticalPlot}, where the maximum velocity is around 4 m/s and the minimum velocity is around -3 m/s suggesting that the roll-up is not fully completed and the vortex is still developing. The measurement section has been selected close to the tip intentionally, as it would be cheaper to conduct numerical simulations of tip vortex flows at closer sections.

\noindent
Similar to previous studies ~\cite{PenningsThesis,Arndt1992,Arndt1991}, an accelerated axial velocity at the vortex core, as high as 1.85 $\text{U}_{\text{inlet}}$, is observed. The axial velocity distribution, Fig. \ref{fig::SmoothAxialPlot}, also suggests that the measurements have insufficient spatial resolution to capture such a high velocity gradient especially in the vortex core region. 

\begin{figure}
		  \centering
		  \selectcolormodel{gray}
       
			\begin{tikzpicture}
			  \begin{axis}[legend pos=south east, xlabel=r/$\text{r}_{\text{vortex}}$,ylabel=Normalized azimuthal velocity, width=0.75\textwidth, height = 0.5\textwidth, xmin=0, xmax=5, ymin=0, ymax=1.1,label style={font=\large},legend cell align={left},legend style={font=\small}]
			  
			  \pgfplotsset{every x tick label/.append style={font=\large}}
			  \pgfplotsset{every y tick label/.append style={font=\large}}		     
			  
				\addplot [very thick, black, mark=square,		mark size=2.0] plot[only marks] file 	{PLOT/AzimuthalVel/SmoothFoil.dat};
				\addplot [very thick, black, mark=triangle,	mark size=2.0] plot[only marks] file 	{PLOT/AzimuthalVel/RoughS40.dat};
				\addplot [very thick, black, mark=o,		mark size=2.0] plot[only marks] file 	{PLOT/AzimuthalVel/Pennings.dat};
				\addplot [very thick, black, mark=+,	mark size=2.0] plot[only marks] file 	{PLOT/AzimuthalVel/Pennings_sigma26.dat};
				\addplot [very thick, black, mark=*,	mark size=1.5] plot[only marks] file 	{PLOT/AzimuthalVel/Arndt.dat};

				\addlegendentry{SmoothFoil, non-cavitating, Re=$1.204 \times 10^6$}
				\addlegendentry{RoughFoil, $\sigma=2.6$, Re=$1.204 \times 10^6$}
				\addlegendentry{SmoothFoil Pennings ~\cite{PenningsThesis}, non-cavitating, Re=$0.9 \times 10^6$}		
				\addlegendentry{SmoothFoil Pennings ~\cite{PenningsThesis}, $\sigma=2.6$, Re=$0.9 \times 10^6$}
				\addlegendentry{SmoothFoil Arndt et al. ~\cite{Arndt1991},  non-cavitating, Re=$0.52 \times 10^6$}
				
			  \end{axis}
			\end{tikzpicture}
		\caption{Variation of the normalized azimuthal velocity in the vortex core region. In each condition, the radial distance is normalized by the vortex core radius and the velocity is normalized by the free-stream velocity. The cavitating condition is reported for $\sigma=2.6$.}
		\label{fig:normalizedAzimuthalVelocityCompar}	
\end{figure}
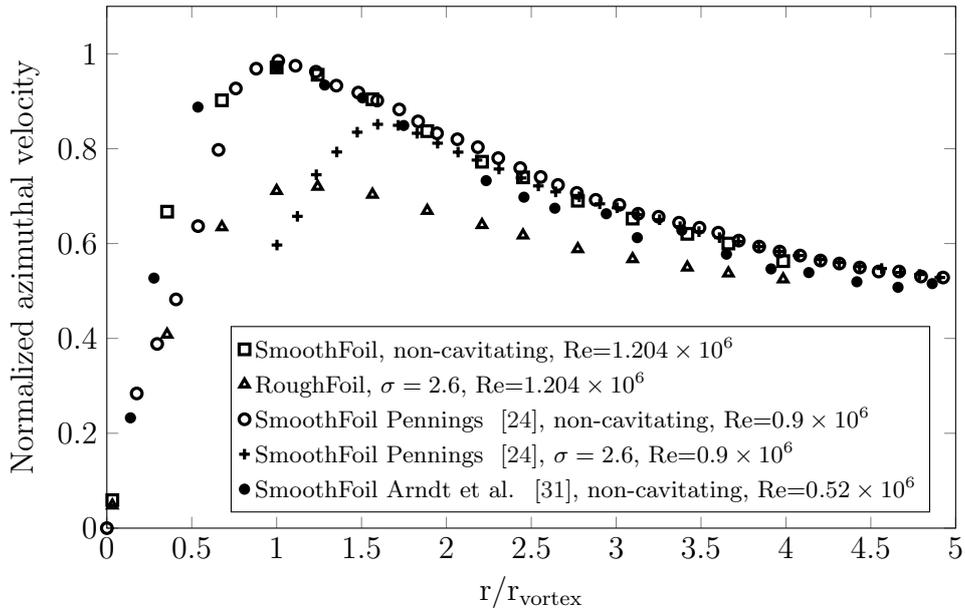

\noindent
The variation of the normalized azimuthal velocity against the radial distance is presented in Fig. \ref{fig:normalizedAzimuthalVelocityCompar}. The velocity profile is plotted over the horizontal cutline, i.e. x-direction, on the section of the LDV measurement which passes through the vortex center. Along with the current measurements, similar measurements on this foil are provided in this figure where the measurements reported by Pennings ~\cite{PenningsThesis} is conducted at $z/\text{C}_0$=1.14 and at Re=$0.9 \times 10^6$ and the measurements reported by Arndt et al. ~\cite{Arndt1991} is conducted at $z/\text{C}_0$=1.0 and at Re=$0.52 \times 10^6$. The measurements include both non-cavitating condition and cavitating condition of $\sigma=2.6$. For the rough foil tests, the roughness pattern (II) with uniform sand grain distribution is reported. In each measurement, the azimuthal velocity is normalized by the inlet velocity and the radial distance is normalized by the vortex core radius. The vortex core radius is considered as the location where the maximum azimuthal velocity occurs, and the vortex core center is the location where the azimuthal velocity is zero. The vortex core radius measured in this study, by Arndt et al. ~\cite{Arndt1991}, and by Pennings ~\cite{PenningsThesis} are $\text{r}_{\text{vortex}}$=1.24, 1.11, 1.1 mm, respectively. 

\noindent
Even though the measurements are conducted at different Reynolds number ranging from $0.52 \times 10^6$ to $1.204 \times 10^6$, a similar trend is observed for the  azimuthal velocity at the non-cavitating condition of the smooth foil, especially at r/$\text{r}_{\text{vortex}}>$1. However, inside the vortex core, r/$\text{r}_{\text{vortex}}<$1, the difference in measurements is more distinguishable. It is not clear whether this difference is a result of having different measurement methods, PIV in Pennings ~\cite{PenningsThesis}, LDV in the current tests and in Arndt et al. ~\cite{Arndt1991}, or having different measurement resolutions. According to the authors knowledge, in neither of the presented data the effect of the tip vortex meandering is accounted.

\noindent
As expected, the presence of cavitation inside the vortex core increases the radius and decreases the maximum azimuthal velocity. As can be seen from results of Pennings at $\sigma=2.6$, the radius of the vaporous tip vortex is around 60 $\%$ larger than the non-cavitating tip vortex. It is interesting that outside the vortex core, the trend of the velocity profiles in cavitating and non-cavitating conditions are similar, suggesting a very little impact of cavitation presence on the vortex roll-up. Having roughness on the foil, however, not only changes the maximum azimuthal velocity, but also affects the vortex roll-up and consequently the trend of the azimuthal velocity variation in r/$\text{r}_{\text{vortex}}>$1.  

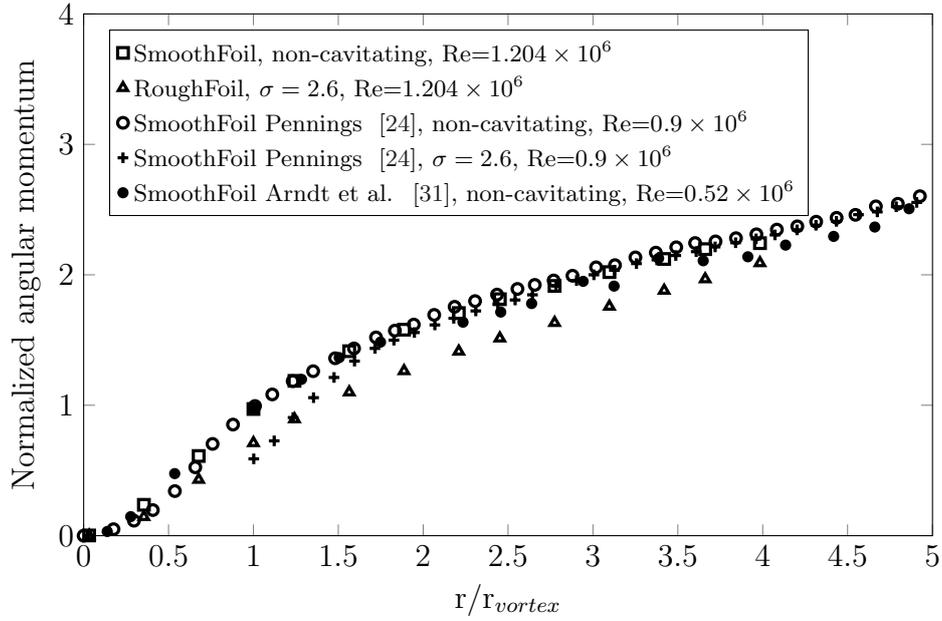
\begin{figure}
		  \centering
		  \selectcolormodel{gray}
       
			\begin{tikzpicture}
			  \begin{axis}[legend pos=north west, xlabel=r/$\text{r}_{vortex}$,ylabel=Normalized angular momentum, width=0.75\textwidth, height = 0.5\textwidth, xmin=0, xmax=5, ymin=0, ymax=4.0,label style={font=\large},legend cell align={left},legend style={font=\small}]
			  
			  \pgfplotsset{every x tick label/.append style={font=\large}}
			  \pgfplotsset{every y tick label/.append style={font=\large}}		     
			  
				\addplot [very thick, black, mark=square,		mark size=2.0] plot[only marks] file 		{PLOT/AngularMomentum/SmoothFoil.dat};
				\addplot [very thick, black, mark=triangle,	mark size=2.0] plot[only marks] file 		{PLOT/AngularMomentum/RoughS40.dat};
				\addplot [very thick, black, mark=o,		mark size=2.0] plot[only marks] file 		{PLOT/AngularMomentum/Pennings.dat};
				\addplot [very thick, black, mark=+,	mark size=2.0] plot[only marks] file 		{PLOT/AngularMomentum/Pennings_sigma26.dat};
				\addplot [very thick, black, mark=*,		mark size=1.5] plot[only marks] file 		{PLOT/AngularMomentum/Arndt.dat};

				\addlegendentry{SmoothFoil, non-cavitating, Re=$1.204 \times 10^6$}
				\addlegendentry{RoughFoil, $\sigma=2.6$, Re=$1.204 \times 10^6$}
				\addlegendentry{SmoothFoil Pennings ~\cite{PenningsThesis}, non-cavitating, Re=$0.9 \times 10^6$}					
				\addlegendentry{SmoothFoil Pennings ~\cite{PenningsThesis}, $\sigma=2.6$, Re=$0.9 \times 10^6$}
				\addlegendentry{SmoothFoil Arndt et al. ~\cite{Arndt1991},  non-cavitating, Re=$0.52 \times 10^6$}				

			  \end{axis}
			\end{tikzpicture}
		\caption{Variation of the normalized angular momentum in the vortex core region.}
		\label{fig:normalizedAngularMomentum}	
\end{figure}

\noindent
By changing the distribution of pressure and boundary layers on the foil, roughness affects the momentum distribution fed into the tip vortex. This can be observed from Fig. \ref{fig:normalizedAngularMomentum} where the distribution of the angular momentum obtained from the product of the normalized azimuthal velocity and the normalized radial distance is presented. Comparison of the angular momentum distributions between different conditions shows that the normalized angular momentum is relatively similar in non-cavitating smooth conditions for the range of tested Reynolds numbers. It is distinguishable that the presence of roughness clearly decreases the angular momentum especially in the vortex core region, i.e. r/$\text{r}_{\text{vortex}}<$5.

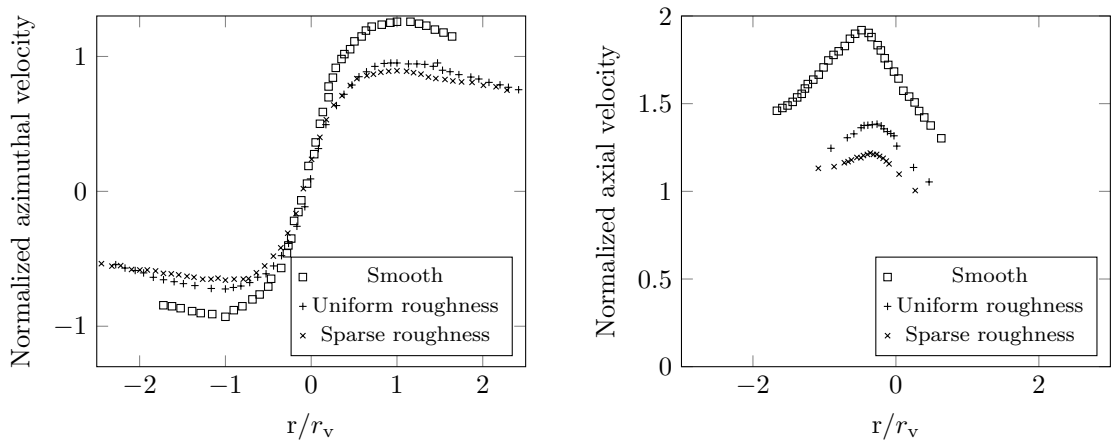
\begin{figure}[h!]
		\centering
		\begin{subfigure}[b]{0.425\textwidth}
			\centering
				  \selectcolormodel{gray}
					\begin{tikzpicture}
					  \begin{axis}[legend pos=south east, legend style={font=\fontsize{8}{5}\selectfont}, xlabel=r/$r_\textnormal{v}$, ylabel=Normalized azimuthal velocity, width=\textwidth, xmin=-2.5, xmax=2.5, ymin=-1.3, ymax=1.3]

						\addplot [black, mark=square,mark size=1.4] plot[only marks] file {PLOT/azimuthalVelNonCavZC05/Exp_smooth.dat};
						\addlegendentry{Smooth}
												
						\addplot [black, mark=+,mark size=1.4] plot[only marks] file 	{PLOT/azimuthalVelNonCavZC05/Exp_s40.dat};
						\addlegendentry{Uniform roughness}								

						\addplot [black, mark=x,mark size=1.4] plot[only marks] file 	{PLOT/azimuthalVelNonCavZC05/Exp_s100Sparse.dat};
						\addlegendentry{Sparse roughness}	
									
					  \end{axis}
					\end{tikzpicture}	
					    \label{fig:normalizedAzimuthalVelocity}	
        \end{subfigure} 
        \quad
		\begin{subfigure}[b]{0.425\textwidth}
			\centering
				  \selectcolormodel{gray}
					\begin{tikzpicture}
					  \begin{axis}[legend pos=south east, legend style={font=\fontsize{8}{5}\selectfont}, xlabel=r/$r_\textnormal{v}$, ylabel=Normalized axial velocity, width=\textwidth, xmin=-3, xmax=3, ymin=0, ymax=2]

						\addplot [black, mark=square,mark size=1.4] plot[only marks] file {PLOT/axialNonCavZC05/Exp_smooth.dat};
						\addlegendentry{Smooth}
												
						\addplot [black, mark=+,mark size=1.4] 		plot[only marks] file {PLOT/axialNonCavZC05/Exp_s40.dat};
						\addlegendentry{Uniform roughness}								
						
						\addplot [black, mark=x,mark size=1.4] plot[only marks] file 	{PLOT/axialNonCavZC05/Exp_s100Sparse.dat};
						\addlegendentry{Sparse roughness}	
				
					  \end{axis}
					\end{tikzpicture}
		    \label{fig:normalizedAxialVelocity}
        \end{subfigure}         
        
		\caption{Variation of the normalized azimuthal and streamwise velocity in the vortex core region at z/$\text{C}_0$=0.5 downstream of the foil tip for the smooth and roughness pattern (II) conditions, non-cavitating, Re=$1.204 \times 10^6$.}
    \label{fig:normalizedAxialAzimuthalVelocity}
\end{figure}

\noindent
The comparison between streamwise and tangential velocity distributions for the smooth and pattern (II) conditions at z/$\text{C}_0$=0.5 downstream of the foil tip is presented in Fig. \ref{fig:normalizedAxialAzimuthalVelocity}. In each condition, the velocity is normalized by using the free-stream velocity, i.e. $\text{U}_\text{inlet}$. By assuming that the vortex core center have zero rotational velocity, half of the distance between maximum and minimum vertical velocities is considered as the vortex radius which then is used in these figures for normalizations. The obtained vortex core radius for the smooth foil is r$_v$=1.24 mm, for the uniform roughness pattern is r$_v$=1.72 mm and for the sparse roughness pattern is r$_v$=1.86 mm, respectively.

\noindent
The measurements indicate that in the roughened surface condition the magnitudes of the tangential and axial velocities are decreased compared to the smooth condition. It is also noted that the axial velocity peak happens slightly off the center, e.g. in the smooth condition it happens at r/r$_v$=0.5. 

\noindent
The sparse roughness distribution leads to slightly lower tangential velocity compared to the uniform roughness distribution. The impact on the axial velocity is more noticeable where lower velocity magnitude is observed in the sparse distribution. Having a lower maximum axial velocity and larger vortex core radius in the sparse roughness pattern indicate that the tip vortex in this pattern is weaker, and consequently it is expected that tip vortex of this pattern has a higher vortex core pressure. 

\subsection{Tip vortex cavitation inception}
 \noindent
The inception point is determined through visual observations of cavitation appearance in the tip vortex region obtained from the high-speed recordings. The condition where the expanded bubble inside the tip vortex can be detected by a naked eye is considered as an inception point. The inception point is selected while lowering the cavitation number from the atmospheric pressure and the impact of hysteresis, like the study by Amini et al. ~\cite{Amini2019Hysteresis,AminiCav2018}, is not included here.

\begin{figure}[h!]
    \centering
		\begin{subfigure}[b]{0.5\textwidth}
		  \centering
		  \selectcolormodel{gray}
			\begin{tikzpicture}
			  \begin{axis}[font=\large,legend pos=north west, xlabel=$\text{C}_\text{l}$, ylabel=$\sigma_i$, width=\textwidth,  height=1.3\textwidth,
         xmin=0, xmax=1.2, ymin=0, ymax=9, xtick={0,0.2,0.4,0.6,0.8,1.0,1.2}, ytick={0,1,2,3,4,5,6,7,8}]
			  
			  \pgfplotsset{every x tick label/.append style={font=\large, yshift=-0.25ex}}
			  \pgfplotsset{every y tick label/.append style={font=\large, xshift=-0.25ex}}

			  	\addplot[solid,mark=triangle*,mark size=2] table[x index=0,y index=1]  {PLOT/SmoothFoil/ExpWeakWaterInception.dat};				
			  	\addplot[only marks, mark=otimes*,error bars/.cd, y dir=both,y explicit]
						 coordinates { 
						(0.4280136022,1.9337231969)+-	(0.6549707602,0.6549707602)
						(0.5913381421,2.5886939571)+-	(0.4054580897,0.4054580897)
						(0.6869592488,2.0584795322)+-	(1.1228070175,1.1228070175)
						(0.7548225629,3.4307992203)+-	(0.5458089669,0.5458089669) };

			  	\addplot[only marks,,mark=square,mark size=2] coordinates { (0.63,6.0)(0.86,7.8) };
						
				\addlegendentry{\footnotesize Weak water}
				\addlegendentry{\footnotesize Strong water}   				      
				\addlegendentry{\footnotesize Current measurements}   				      																								
			  \end{axis}
			\end{tikzpicture}
		\end{subfigure}				
    \caption{Variation of the cavitation inception and lift coefficient, smooth foil, Re=$8.56 \times 10^5$, weak and strong water data are extracted from Arndt and Keller ~\cite{Arndt1992}.}
    \label{fig:SigmaClVariationSmooth}
\end{figure}
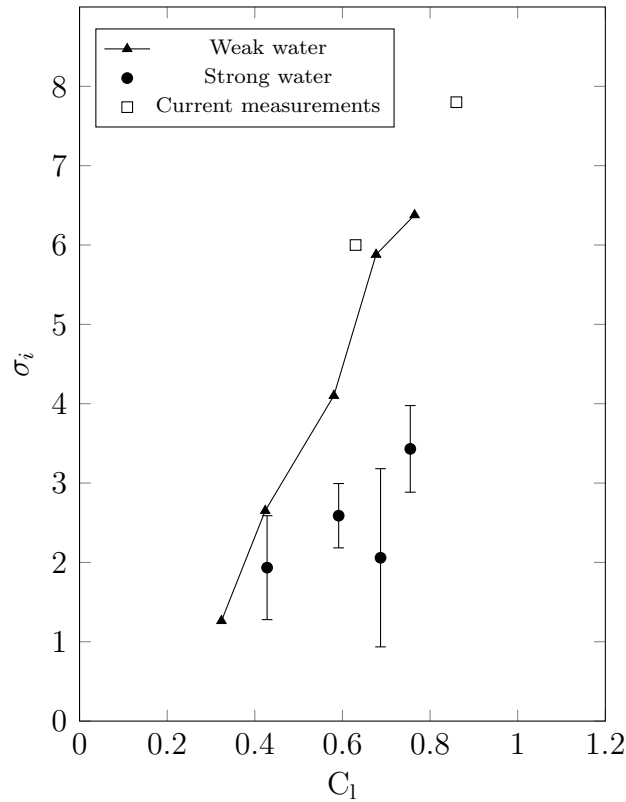

\noindent
During the tests, fresh water was added to the cavitation tunnel upstream of the foil to provide a similar water quality. This leads to having relatively weak water quality for all of the presented measured data. This can be noted from Fig. \ref{fig:SigmaClVariationSmooth} where the measured inception point for the smooth foil is presented. Comparison with the measurements conducted by Arndt and Keller ~\cite{Arndt1992} where different water qualities were tested indicates that the water quality of the current tests can be considered weak, i.e. filled with enough nuclei to initiate cavitation as soon as pressure falls below the saturation pressure. It can be noted that the measured inception points are in good agreement with the data provided in ~\cite{Arndt1992}.  

\begin{figure}[h!]
	\centering  
			  \selectcolormodel{gray}
    \begin{tikzpicture}
        \begin{axis}[
            symbolic x coords={Smooth, (I), (II)-Uniform, (II)-Sparse},
            xtick=data,	ylabel={$\sigma_i$}, xlabel={Roughness pattern},   y label style={at={(-0.1,0.5)}},
             ymin=0, ymax=8, ybar,bar width = {1.2em}, width=8.8 cm, height=7 cm]			          
            \addplot[pattern=north east lines]  coordinates {
                (Smooth,       					6.0)
                ((I),   						4.15)
                ((II)-Uniform,   					4.05)
                ((II)-Sparse,   					3.3)                
            };		            
        \end{axis}
    \end{tikzpicture}
    
    \caption{Variation of the cavitation inception versus different surface roughness areas, Re=$8.55 \times 10^5$, $\alpha$= 9$^\circ$.}
    \label{fig:varCavInceptRoughArea}    
\end{figure}
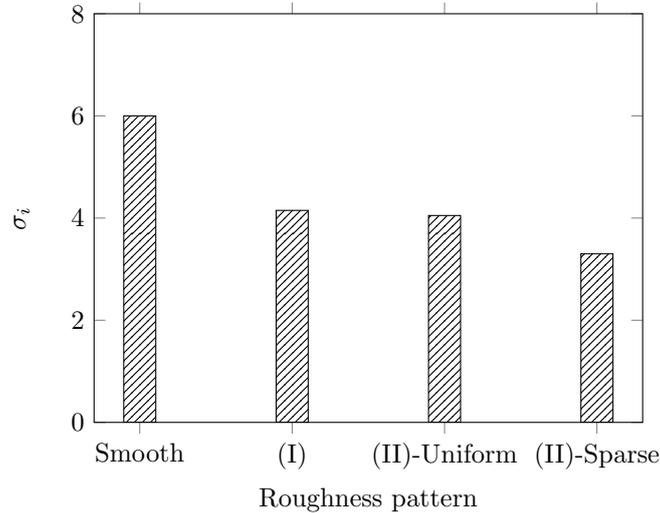

\noindent
The predicted cavitation inception points for different roughness patterns are presented in Fig. \ref{fig:varCavInceptRoughArea}. For the tested roughness patterns, application of roughness has mitigated the TVCI by around 33$\%$. The predicted inception point for pattern (I) is $\sigma_i=4.15$ and for pattern (II) with uniform sand distribution is $\sigma_i=4.05$ and with the sparse distribution is $\sigma_i=3.3$. No difference in the cavitation inception properties is observed while comparing the uniform and non-uniform roughness distributions. This, however, does not clarify whether it is the boundary layer thickness increase or the turbulent instabilities which is the main responsible for TVC mitigation by roughness.

\noindent
It is already discussed that the tip vortex pressure of the sparse pattern is expected to be higher than that of the uniform roughness pattern. This corresponds to a lower cavitation inception in the sparse pattern compared to the uniform pattern.

\noindent
At the inception point of different roughness patterns, similar lift coefficient is measured where the lift coefficient variation is less than 0.2 $\%$. The drag coefficient of the roughness patterns are higher than the smooth foil, increased by 3 $\%$ in the pattern (I) and by 3.5 $\%$ in the pattern (II).
 	
\subsection{Cavitating tip vortex}
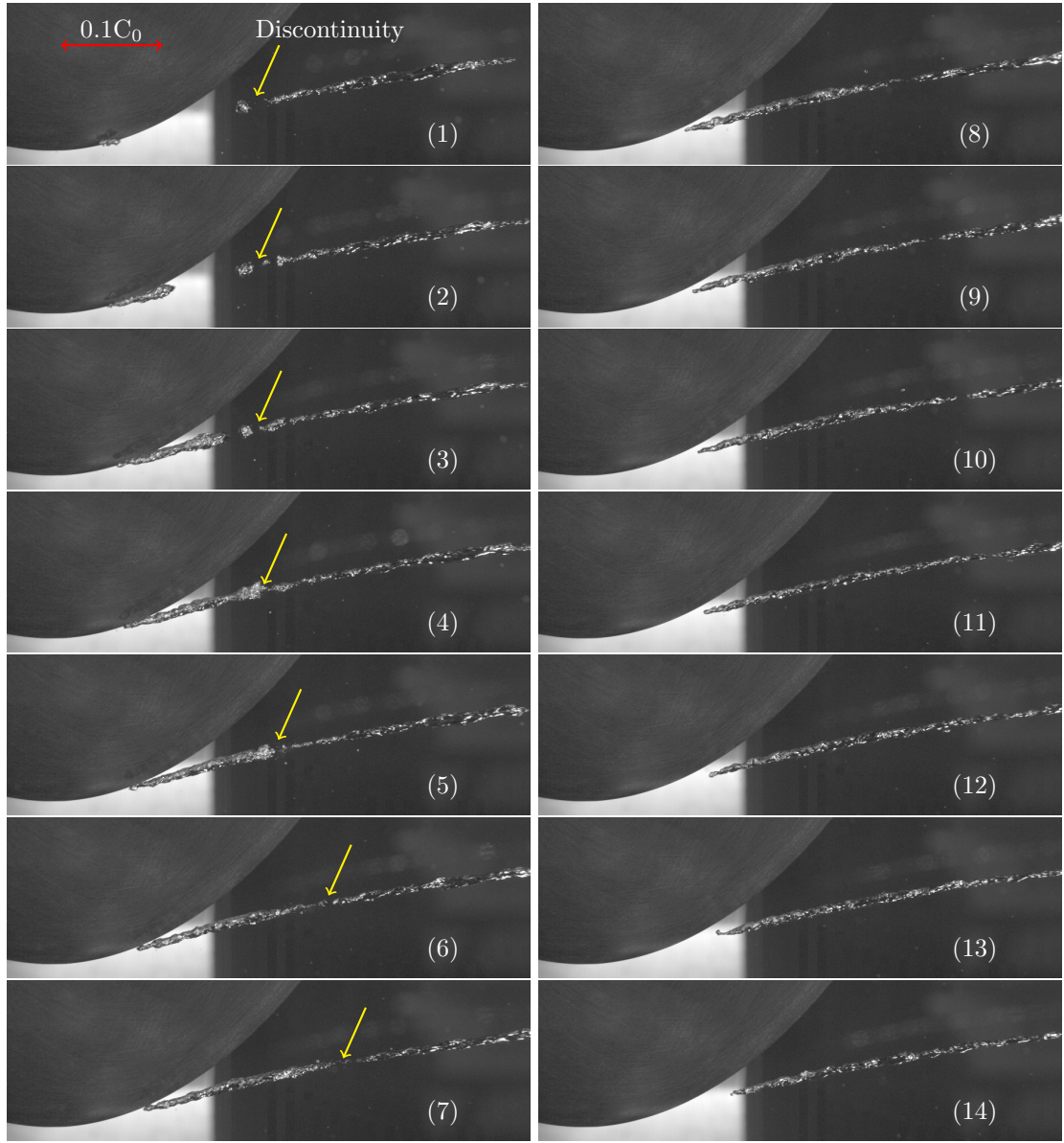
\begin{figure} [h!]
        \centering    
        \begin{subfigure}[b]{0.423\textwidth}
				  \begin{tikzpicture}
		\node [anchor=south west,inner sep=0](mesh)at(0,0){\includegraphics[width=\textwidth]{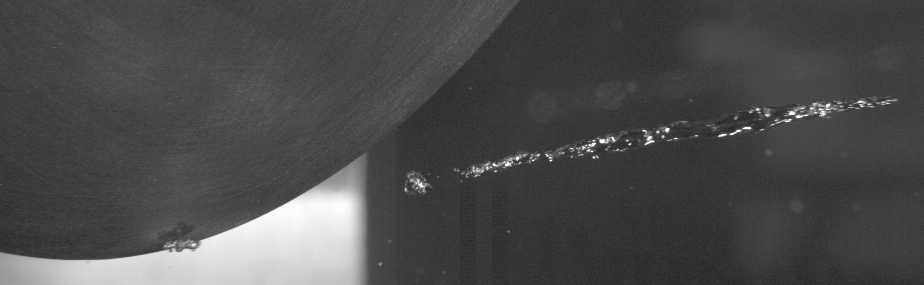}};
					\node [draw=none, color =white] at (6.0,0.4) {(1)}; 		
					\draw [<->, thick, color =red] (0.75,1.65) -- (2.15,1.65);	
					\node [draw=none, color =white] at (1.45,1.85) {0.1$\text{C}_0$}; 	
					
					\draw [->, thick, color =yellow] (3.75,1.65) -- (3.44,0.95);	
					\node [draw=none, color =white] at (4.45,1.85) {Discontinuity}; 																							
				  \end{tikzpicture}	  
        \end{subfigure}
         \begin{subfigure}[b]{0.423\textwidth}
				  \begin{tikzpicture}
		\node [anchor=south west,inner sep=0](mesh)at(0,0){\includegraphics[width=\textwidth]{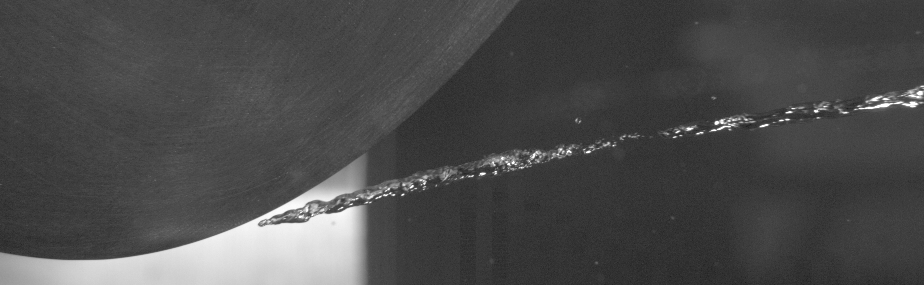}};
					\node [draw=none, color =white] at (6.0,0.4) {(8)}; 	
				  \end{tikzpicture}	   
        \end{subfigure}      
        \begin{subfigure}[b]{0.423\textwidth}
				  \begin{tikzpicture}
		\node [anchor=south west,inner sep=0](mesh)at(0,0){\includegraphics[width=\textwidth]{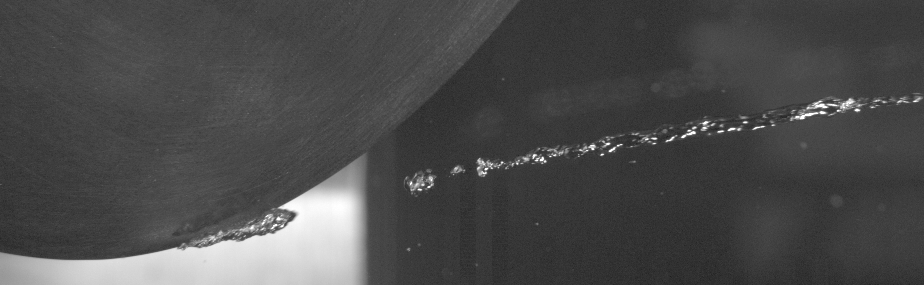}};
					\node [draw=none, color =white] at (6.0,0.4) {(2)}; 	
					\draw [->, thick, color =yellow] (3.78,1.65) -- (3.47,0.95);																	   
				  \end{tikzpicture}	   
        \end{subfigure}
         \begin{subfigure}[b]{0.423\textwidth}
				  \begin{tikzpicture}
		\node [anchor=south west,inner sep=0](mesh)at(0,0){\includegraphics[width=\textwidth]{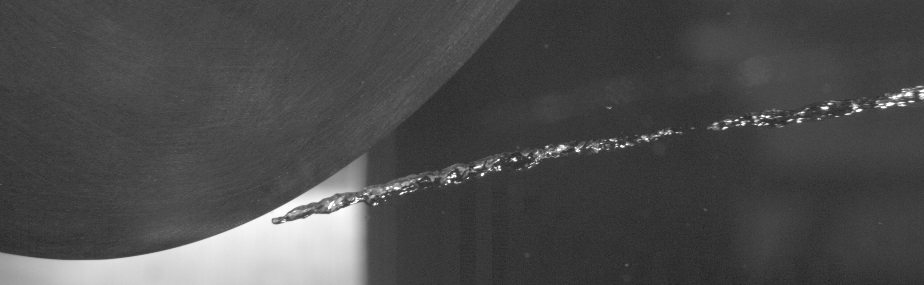}};
					\node [draw=none, color =white] at (6.0,0.4) {(9)};  								
				  \end{tikzpicture}	   
        \end{subfigure}    
        \begin{subfigure}[b]{0.423\textwidth}
				  \begin{tikzpicture}
		\node [anchor=south west,inner sep=0](mesh)at(0,0){\includegraphics[width=\textwidth]{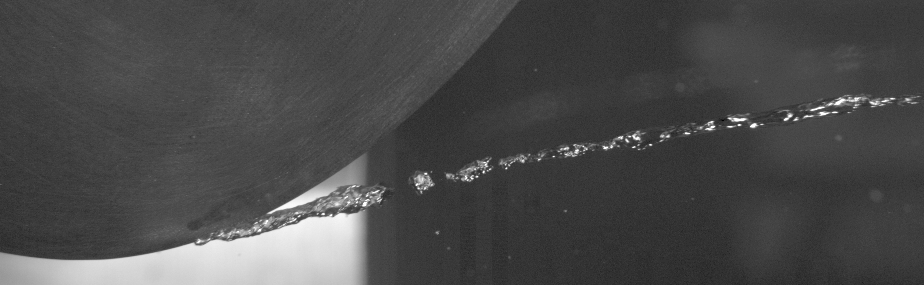}};
					\node [draw=none, color =white] at (6.0,0.4) {(3)}; 
					\draw [->, thick, color =yellow] (3.78,1.65) -- (3.47,0.95);	 								
				  \end{tikzpicture}	   
        \end{subfigure}
         \begin{subfigure}[b]{0.423\textwidth}
				  \begin{tikzpicture}
		\node [anchor=south west,inner sep=0](mesh)at(0,0){\includegraphics[width=\textwidth]{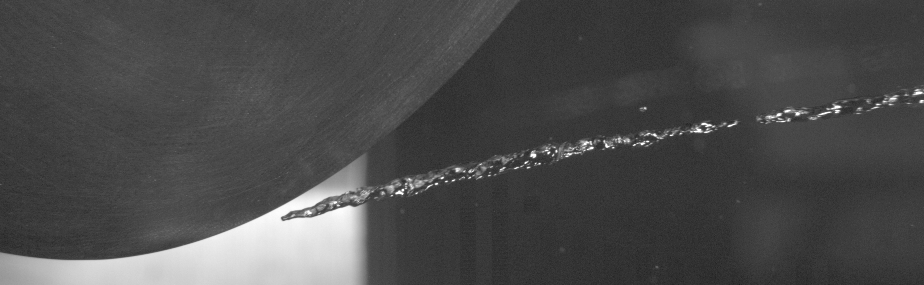}};
					\node [draw=none, color =white] at (6.0,0.4) {(10)};  								
				  \end{tikzpicture}	   
        \end{subfigure}      
         \begin{subfigure}[b]{0.423\textwidth}
				  \begin{tikzpicture}
		\node [anchor=south west,inner sep=0](mesh)at(0,0){\includegraphics[width=\textwidth]{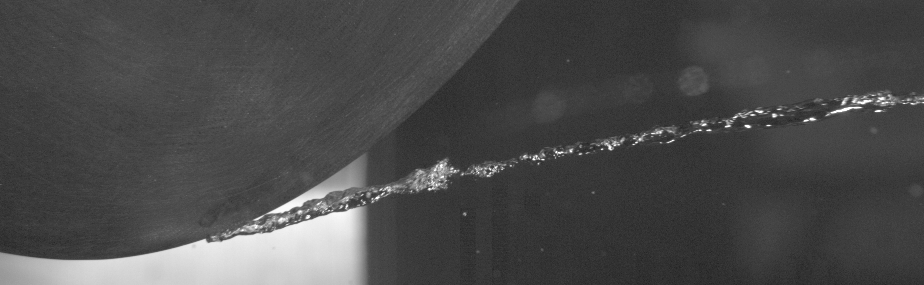}};
					\node [draw=none, color =white] at (6.0,0.4) {(4)}; 
					\draw [->, thick, color =yellow] (3.85,1.65) -- (3.54,0.95);														   								
				  \end{tikzpicture}	   
        \end{subfigure}
         \begin{subfigure}[b]{0.423\textwidth}
				  \begin{tikzpicture}
		\node [anchor=south west,inner sep=0](mesh)at(0,0){\includegraphics[width=\textwidth]{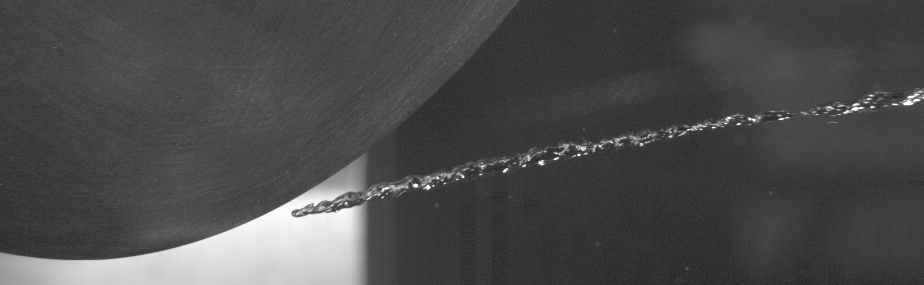}};
					\node [draw=none, color =white] at (6.0,0.4) {(11)}; 												   								
				  \end{tikzpicture}	   
        \end{subfigure}       
         \begin{subfigure}[b]{0.423\textwidth}
				  \begin{tikzpicture}
		\node [anchor=south west,inner sep=0](mesh)at(0,0){\includegraphics[width=\textwidth]{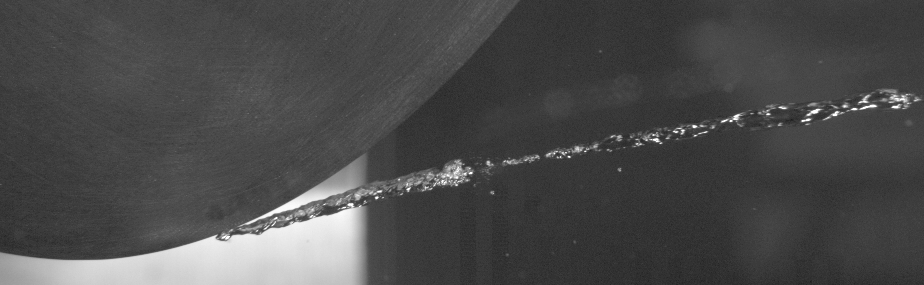}};
					\node [draw=none, color =white] at (6.0,0.4) {(5)}; 
										\draw [->, thick, color =yellow]  (4.05,1.75) -- (3.74,1.05);					   								
				  \end{tikzpicture}	   
        \end{subfigure}
         \begin{subfigure}[b]{0.423\textwidth}
				  \begin{tikzpicture}
		\node [anchor=south west,inner sep=0](mesh)at(0,0){\includegraphics[width=\textwidth]{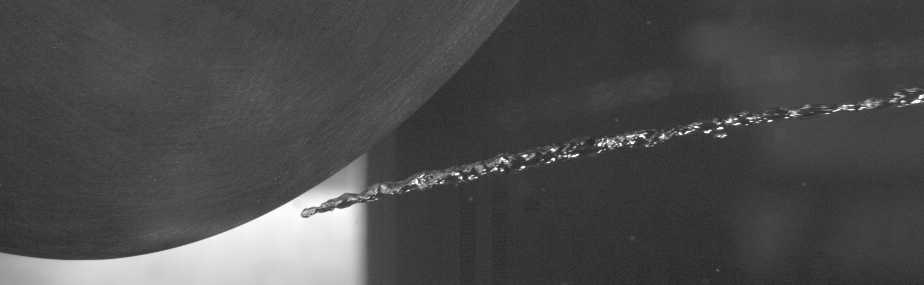}};
					\node [draw=none, color =white] at (6.0,0.4) {(12)}; 												   								
				  \end{tikzpicture}	   
        \end{subfigure}        
         \begin{subfigure}[b]{0.423\textwidth}
				  \begin{tikzpicture}
		\node [anchor=south west,inner sep=0](mesh)at(0,0){\includegraphics[width=\textwidth]{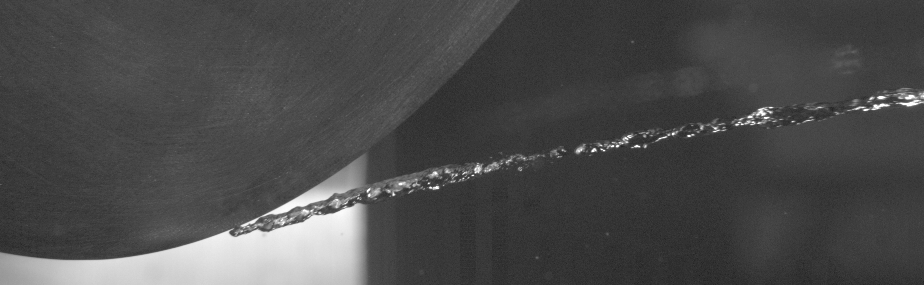}};
					\node [draw=none, color =white] at (6.0,0.4) {(6)}; 
										\draw [->, thick, color =yellow] (4.74,1.85) -- (4.43,1.15);									
				  \end{tikzpicture}	   
        \end{subfigure}
         \begin{subfigure}[b]{0.423\textwidth}
				  \begin{tikzpicture}
		\node [anchor=south west,inner sep=0](mesh)at(0,0){\includegraphics[width=\textwidth]{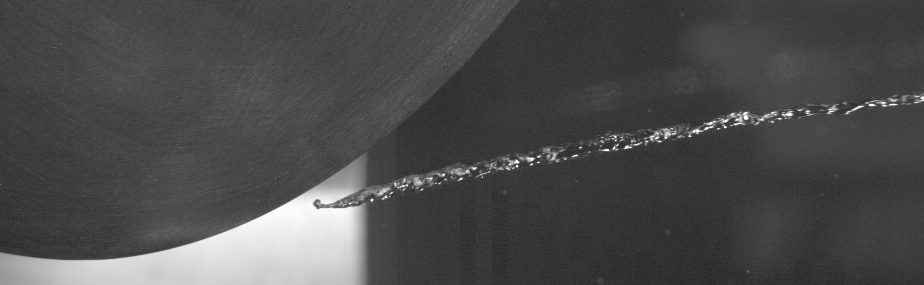}};
					\node [draw=none, color =white] at (6.0,0.4) {(13)}; 												   								
				  \end{tikzpicture}	   
        \end{subfigure}        
         \begin{subfigure}[b]{0.423\textwidth}
				  \begin{tikzpicture}
		\node [anchor=south west,inner sep=0](mesh)at(0,0){\includegraphics[width=\textwidth]{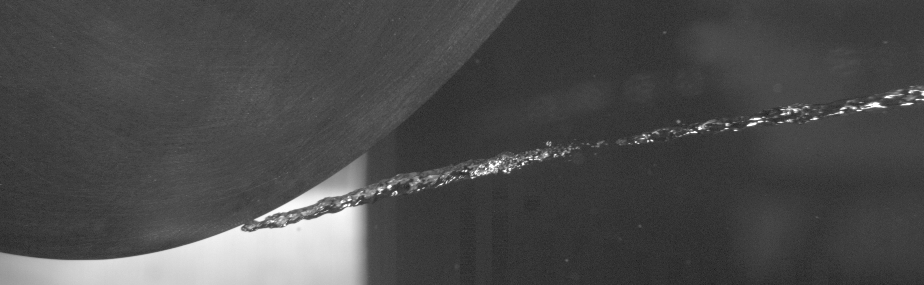}};
					\node [draw=none, color =white] at (6.0,0.4) {(7)}; 
										\draw [->, thick, color =yellow](4.94,1.85) -- (4.63,1.15);										
				  \end{tikzpicture}	   
        \end{subfigure}
         \begin{subfigure}[b]{0.423\textwidth}
				  \begin{tikzpicture}
		\node [anchor=south west,inner sep=0](mesh)at(0,0){\includegraphics[width=\textwidth]{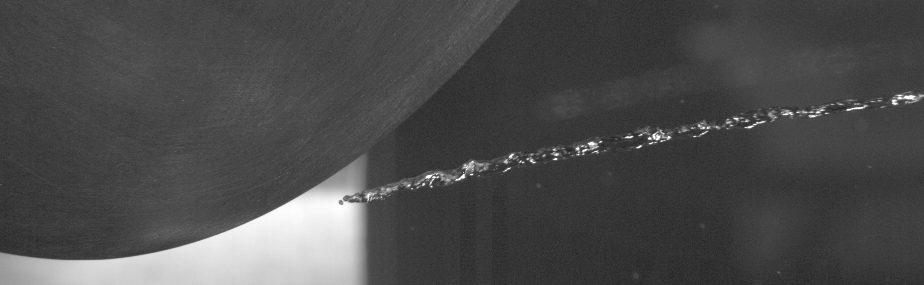}};
					\node [draw=none, color =white] at (6.0,0.4) {(14)}; 												   								
				  \end{tikzpicture}	   
        \end{subfigure}
		\caption{Cavitating tip vortex flow around the smooth foil, $\sigma=2.6$, Re=$8.55 \times 10^5$, $\alpha$= 9$^\circ$. The time step between the consequent images is 1.6 ms.}
		\label{fig::smoothCavTVTransient}
\end{figure}

 \noindent
For the smooth foil condition, the cavitating tip vortex can be decomposed into two areas. Close to the foil and within the downstream distance of 0.2 $\text{C}_0$ from the tip, an intermittent formation of cavitating vortex is observed while further downstream, a relatively stationary cavitating tip vortex is shaped. At the time instance where there is not any attached cavitation to the foil, a small cavitating vortex forms at the tip, Fig. \ref{fig::smoothCavTVTransient} time instance (1). This cavitating vortex grows and eventually reaches the stationary cavitating tip vortex part, Fig. \ref{fig::smoothCavTVTransient} time instances (1, 2 and 3). It is interesting that the location where the stationary cavitating tip vortex starts is relatively constant during this process, having the approximate location of 0.18 $\text{C}_0$ downstream of the tip. It can be noted that there is a clear discontinuity on the stationary cavitating tip vortex, illustrated by the yellow arrow on the figure. When the growing attached cavity reaches the stationary part, it separates from the foil, and is transported downstream, time instances (7) to (14) of Fig. \ref{fig::smoothCavTVTransient}. The noise measurements conducted by Peng et al. ~\cite{Peng20191170} indicates that this TVC detachment process from the foil tip corresponds to a local peak in the tip vortex generated noise. 

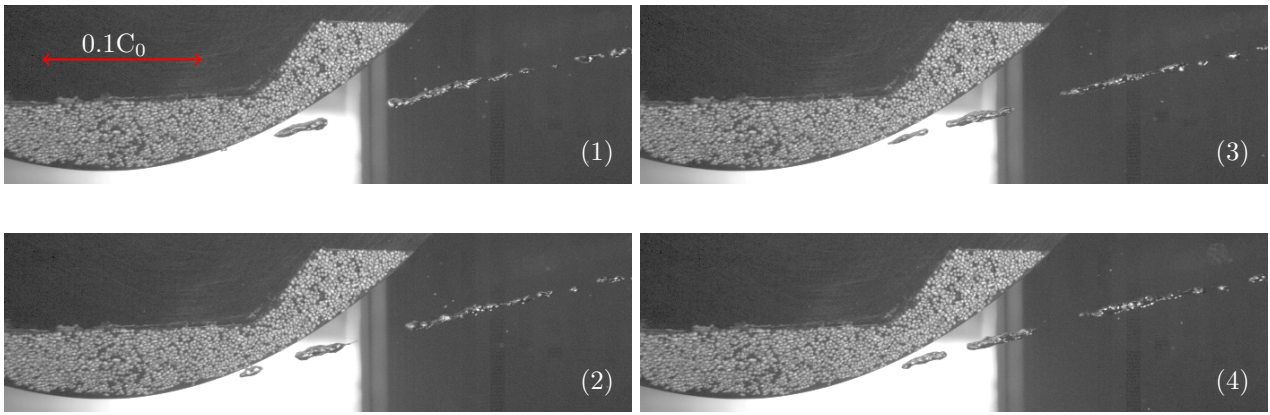
\begin{figure} [h!]
        \centering    
        
        \begin{subfigure}[b]{0.485\textwidth}
				  \begin{tikzpicture}
		\node [anchor=south west,inner sep=0](mesh)at(0,0){\includegraphics[width=\textwidth]{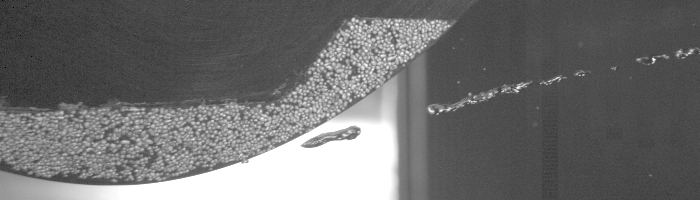}};
					\node [draw=none, color =white] at (7.8,0.4) {(1)}; 	
					\draw [<->, thick, color =red] (0.5,1.65) -- (2.6,1.65);	
					\node [draw=none, color =white] at (1.45,1.85) {0.1$\text{C}_0$}; 							    
				  \end{tikzpicture}	   
                \label{fig::starMesh1}
        \end{subfigure}   
        \begin{subfigure}[b]{0.485\textwidth}
				  \begin{tikzpicture}
		\node [anchor=south west,inner sep=0](mesh)at(0,0){\includegraphics[width=\textwidth]{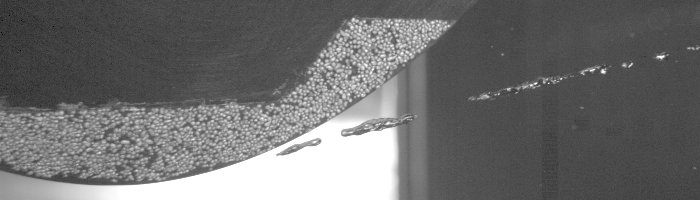}};
					\node [draw=none, color =white] at (7.8,0.4) {(3)}; 			    
				  \end{tikzpicture}	   
                \label{fig::starMesh1}
        \end{subfigure}            
        \begin{subfigure}[b]{0.485\textwidth}
				  \begin{tikzpicture}
		\node [anchor=south west,inner sep=0](mesh)at(0,0){\includegraphics[width=\textwidth]{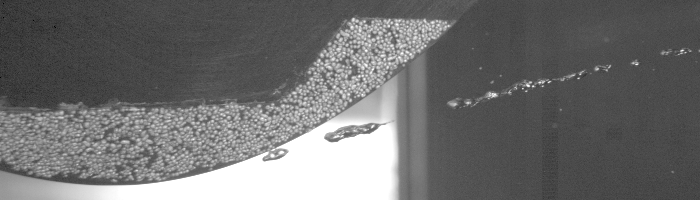}};
					\node [draw=none, color =white] at (7.8,0.4) {(2)}; 			    
				  \end{tikzpicture}	   
                \label{fig::starMesh1}
        \end{subfigure}                
        \begin{subfigure}[b]{0.485\textwidth}
				  \begin{tikzpicture}
		\node [anchor=south west,inner sep=0](mesh)at(0,0){\includegraphics[width=\textwidth]{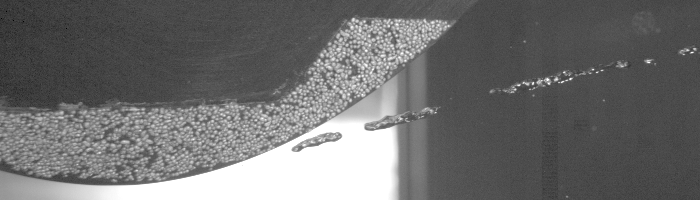}};
					\node [draw=none, color =white] at (7.8,0.4) {(4)}; 			    
				  \end{tikzpicture}	   
                \label{fig::starMesh1}
        \end{subfigure}                                                
		\caption{Cavitating tip vortex around the foil with the roughness pattern (II), $\sigma=2.6$, Re=$8.55 \times 10^5$, $\alpha$= 9$^\circ$. The time step between each image is 1.2 ms.}
		\label{fig::SSLESSTEroughCavTVTransient}
\end{figure}

\noindent
When the roughness is applied on the leading edge and tip of the foil, pattern (II), the cavitating tip vortex becomes much weaker than the one formed on the smooth foil. The cavitating tip vortex is also observed to consist of shed cavitating vortices formed on the tip of the foil, Fig. \ref{fig::SSLESSTEroughCavTVTransient}. Contrary to the smooth foil condition, where a continuous cavitating tip vortex appears after 0.2 $\text{C}_0$, with pattern (II) the discontinuity between shed cavity structures remains as they travel downstream. This can be related to the fact that the tip vortex has become weaker in this condition and therefore cannot create the same cavitating vortex elongation as in the smooth foil condition. It is also noted that formation of cavitating vortex at the tip happens with higher frequency compared to the smooth foil condition.  

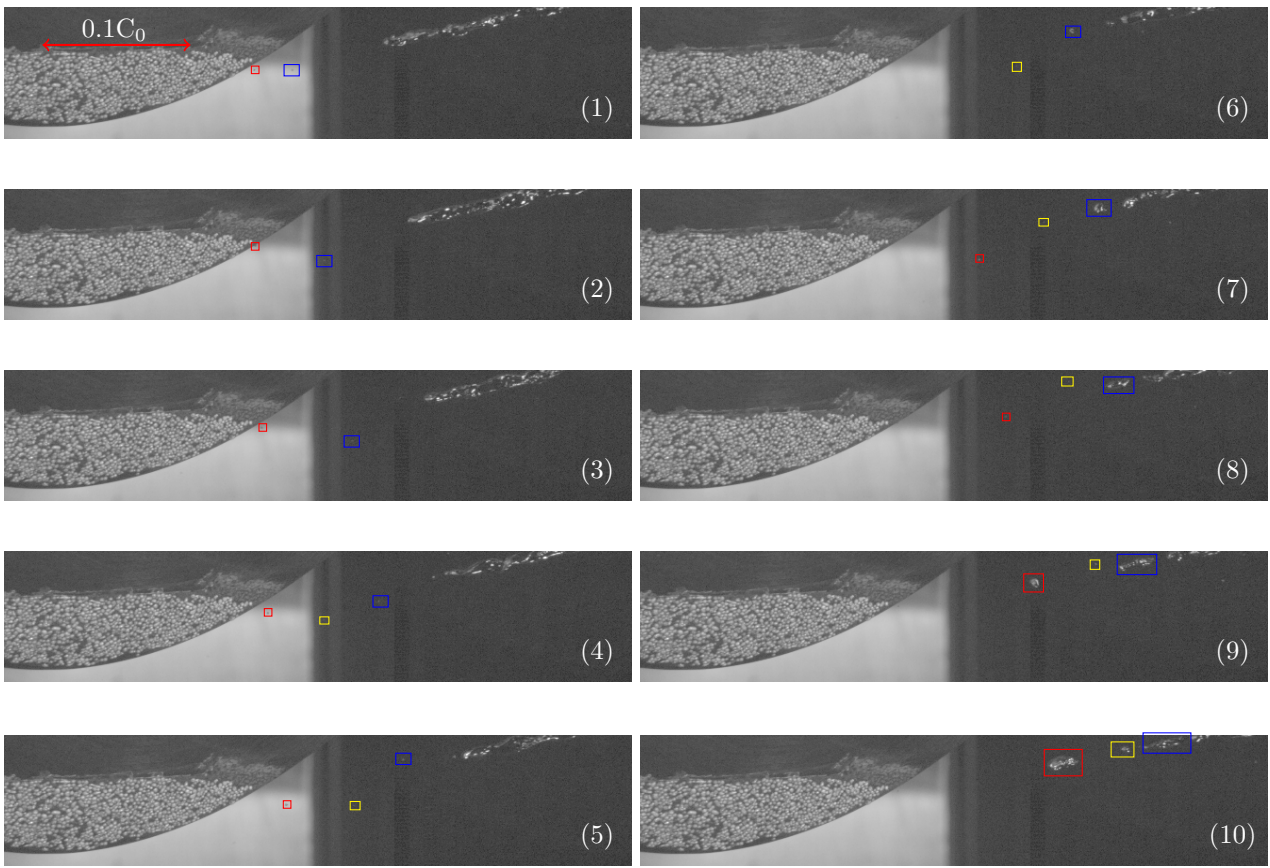
\begin{figure} [h!]
        \centering    
        
        \begin{subfigure}[b]{0.485\textwidth}
				  \begin{tikzpicture}
		\node [anchor=south west,inner sep=0](mesh)at(0,0){\includegraphics[width=\textwidth]{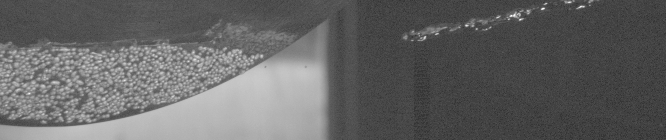}};
					\node [draw=none, color =white] at (7.8,0.4) {(1)}; 
				    \draw[red]  (3.25,0.87) rectangle (3.35,0.97);
				    \draw[blue] (3.68,0.84) rectangle (3.88,0.99);		
					\draw [<->, thick, color =red] (0.5,1.25) -- (2.45,1.25);	
					\node [draw=none, color =white] at (1.45,1.45) {0.1$\text{C}_0$};		    		    
				  \end{tikzpicture}	   
                \label{fig::starMesh1}
        \end{subfigure}
          \begin{subfigure}[b]{0.485\textwidth}
				  \begin{tikzpicture}
		\node [anchor=south west,inner sep=0](mesh)at(0,0){\includegraphics[width=\textwidth]{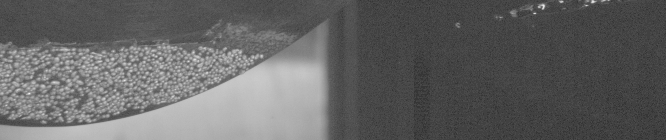}};
					\node [draw=none, color =white] at (7.8,0.4) {(6)}; 
				    \draw[blue] (5.6,1.35) rectangle (5.8,1.5);																	   					
				    \draw[yellow] (4.9,0.9) rectangle (5.02,1.02);																			 
				    								
				  \end{tikzpicture}	   
                \label{fig::starMesh1}
        \end{subfigure}        
        \begin{subfigure}[b]{0.485\textwidth}
				  \begin{tikzpicture}
		\node [anchor=south west,inner sep=0](mesh)at(0,0){\includegraphics[width=\textwidth]{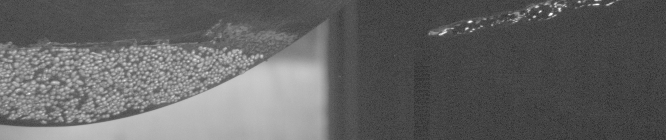}};
					\node [draw=none, color =white] at (7.8,0.4) {(2)}; 	
				    \draw[red]  (3.25,0.93) rectangle (3.35,1.03);
				    \draw[blue] (4.11,0.71) rectangle (4.31,0.86);																  
				  \end{tikzpicture}	   
                \label{fig::starMesh1}
        \end{subfigure}
          \begin{subfigure}[b]{0.485\textwidth}
				  \begin{tikzpicture}
		\node [anchor=south west,inner sep=0](mesh)at(0,0){\includegraphics[width=\textwidth]{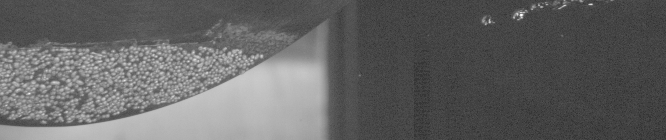}};
					\node [draw=none, color =white] at (7.8,0.4) {(7)}; 
				    \draw[red]  (4.42,0.77) rectangle (4.52,0.87);
				    \draw[blue] (5.88,1.38) rectangle (6.2,1.6);	
					\draw[yellow] (5.25,1.25) rectangle (5.375,1.35);																			  				    																   				    
				  \end{tikzpicture}	   
                \label{fig::starMesh1}
        \end{subfigure}        
          \begin{subfigure}[b]{0.485\textwidth}
				  \begin{tikzpicture}
		\node [anchor=south west,inner sep=0](mesh)at(0,0){\includegraphics[width=\textwidth]{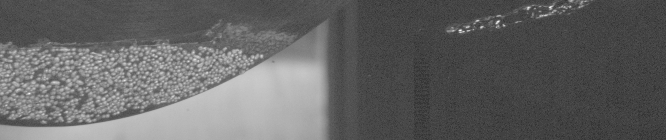}};
					\node [draw=none, color =white] at (7.8,0.4) {(3)}; 
				    \draw[red]  (3.35,0.93) rectangle (3.45,1.03);
				    \draw[blue] (4.47,0.72) rectangle (4.67,0.87);																	   								
				  \end{tikzpicture}	   
                \label{fig::starMesh1}
        \end{subfigure}
          \begin{subfigure}[b]{0.485\textwidth}
				  \begin{tikzpicture}
		\node [anchor=south west,inner sep=0](mesh)at(0,0){\includegraphics[width=\textwidth]{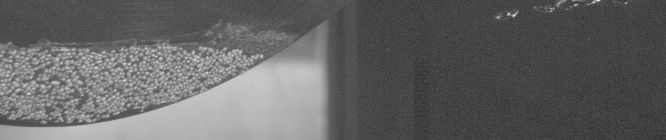}};
					\node [draw=none, color =white] at (7.8,0.4) {(8)}; 
				    \draw[red]  (4.77,1.07) rectangle (4.87,1.17);
				    \draw[blue] (6.1,1.43) rectangle (6.5,1.65);																	  
					\draw[yellow] (5.55,1.53) rectangle (5.7,1.65);																			   								
				  \end{tikzpicture}	   
                \label{fig::starMesh1}
        \end{subfigure}        
          \begin{subfigure}[b]{0.485\textwidth}
				  \begin{tikzpicture}
		\node [anchor=south west,inner sep=0](mesh)at(0,0){\includegraphics[width=\textwidth]{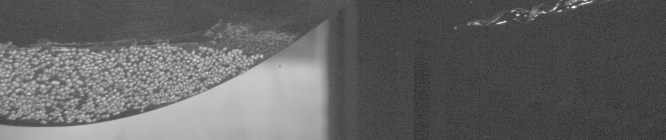}};
					\node [draw=none, color =white] at (7.8,0.4) {(4)}; 
				    \draw[red]  (3.42,0.88) rectangle (3.52,0.98);
				    \draw[blue] (4.85,1.0) rectangle (5.05,1.15);																	   						\draw[yellow] (4.15,0.775) rectangle (4.27,0.87);																			
				  \end{tikzpicture}	   
                \label{fig::starMesh1}
        \end{subfigure}
        \begin{subfigure}[b]{0.485\textwidth}
				  \begin{tikzpicture}
		\node [anchor=south west,inner sep=0](mesh)at(0,0){\includegraphics[width=\textwidth]{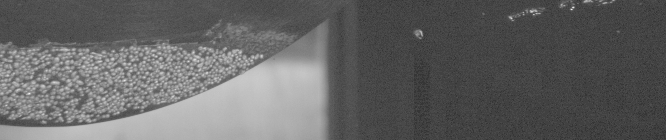}};
					\node [draw=none, color =white] at (7.8,0.4) {(9)}; 
				    \draw[red]  (5.05,1.2) rectangle (5.31,1.44);
				    \draw[blue] (6.28,1.43) rectangle (6.8,1.7);																	   						\draw[yellow] (5.925,1.5) rectangle (6.05,1.625);																
				  \end{tikzpicture}	   
                \label{fig::starMesh1}
        \end{subfigure}        
          \begin{subfigure}[b]{0.485\textwidth}
				  \begin{tikzpicture}
		\node [anchor=south west,inner sep=0](mesh)at(0,0){\includegraphics[width=\textwidth]{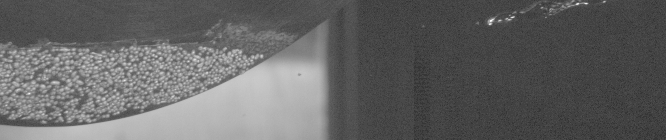}};
					\node [draw=none, color =white] at (7.8,0.4) {(5)}; 
				    \draw[red]  (3.67,0.77) rectangle (3.77,0.87);
				    \draw[blue] (5.15,1.35) rectangle (5.35,1.5);																	   				    					\draw[yellow] (4.55,0.75) rectangle (4.68,0.86);											
				    
				  \end{tikzpicture}	   
                \label{fig::starMesh1}
        \end{subfigure}
        \begin{subfigure}[b]{0.485\textwidth}
				  \begin{tikzpicture}
		\node [anchor=south west,inner sep=0](mesh)at(0,0){\includegraphics[width=\textwidth]{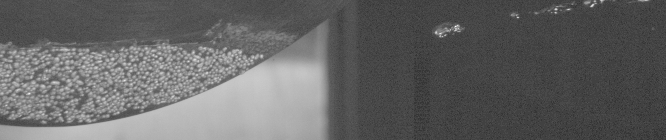}};
					\node [draw=none, color =white] at (7.8,0.4) {(10)};
				    \draw[red]  (5.32,1.2) rectangle (5.82,1.55);
				    \draw[blue] (6.62,1.5) rectangle (7.25,1.77);																	   						\draw[yellow] (6.2,1.45) rectangle (6.5,1.65);								
				  \end{tikzpicture}	   
                \label{fig::starMesh1}
        \end{subfigure}
              
		\caption{Cavitating tip vortex around the foil with the roughness pattern (I), $\sigma=2.6$, Re=$8.55 \times 10^5$, $\alpha$= 9$^\circ$. The time step between each image is 1.6 ms. The colorful rectangles are used to highlight the locations of nuclei entering into the tip vortex.}
		\label{fig::SSLEroughCavTVTransient}
\end{figure}

\noindent
While in the smooth and pattern (II) conditions it is observed that cavitating vortices are shed from the tip and fed into the downstream cavitating tip vortex, with roughness pattern (I) no shedding behaviour close to the tip is observed. In this case, it is found that the cavitating behaviour is more dependent on the distributions of nuclei and their trajectories. In Fig. \ref{fig::SSLEroughCavTVTransient}, three different nuclei having different trajectories are highlighted. As can be seen, depending on the location, that a nucleus enters into the low pressure region of the tip vortex, cavitation forms at different locations. 

\noindent
During the tests and with naked eyes, local formation of very small bubble or sheet cavitation around the roughness elements is observed. The analysis of high-speed recordings shows that these tiny nucleation sites having micro- or even nano-metric residual air pockets are formed among the roughness sands. This, however, does not influence the characteristic of the roughness elements in not inducing a big sheet cavity in the tested operating conditions.

\begin{figure}
        \centering 
        \begin{subfigure}[b]{0.7\textwidth}
				  \begin{tikzpicture}
		\node [anchor=south west,inner sep=0](mesh)at(0,0){\includegraphics[width=\textwidth]{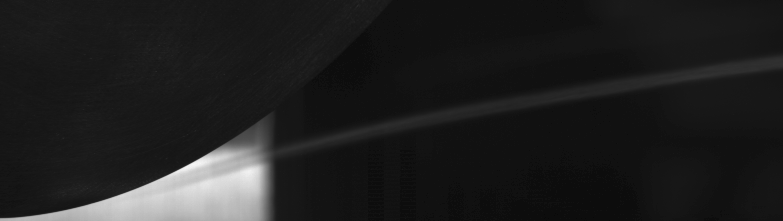}};
					\node [draw=none, color =white] at (6.0,0.4) {Smooth}; 												  	
				  \end{tikzpicture}	   
                \label{fig::smoothTimeAveCTV}
        \end{subfigure}          
        \begin{subfigure}[b]{0.7\textwidth}
				  \begin{tikzpicture}
		\node [anchor=south west,inner sep=0](mesh)at(0,0){\includegraphics[width=\textwidth]{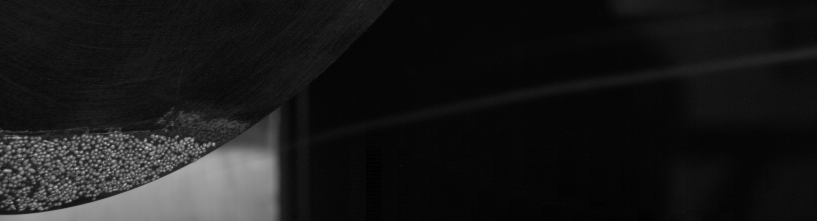}};
					\node [draw=none, color =white] at (6.0,0.4) {Pattern (I)}; 												   								
				  \end{tikzpicture}	   
                \label{fig::smoothTimeAveCTV}
        \end{subfigure}   
        \begin{subfigure}[b]{0.7\textwidth}
				  \begin{tikzpicture}
		\node [anchor=south west,inner sep=0](mesh)at(0,0){\includegraphics[width=\textwidth]{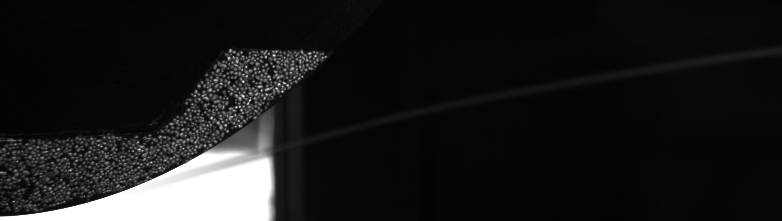}};
					\node [draw=none, color =white] at (6.0,0.4) {Pattern (II)}; 												   								
				  \end{tikzpicture}	   
                \label{fig::smoothTimeAveCTV}
        \end{subfigure}                          
		\caption{Time averaged cavitating tip vortex, $\sigma=2.6$, Re=$8.55 \times 10^5$, $\alpha$= 9$^\circ$.}
		\label{fig::timeAveCTV}
\end{figure}
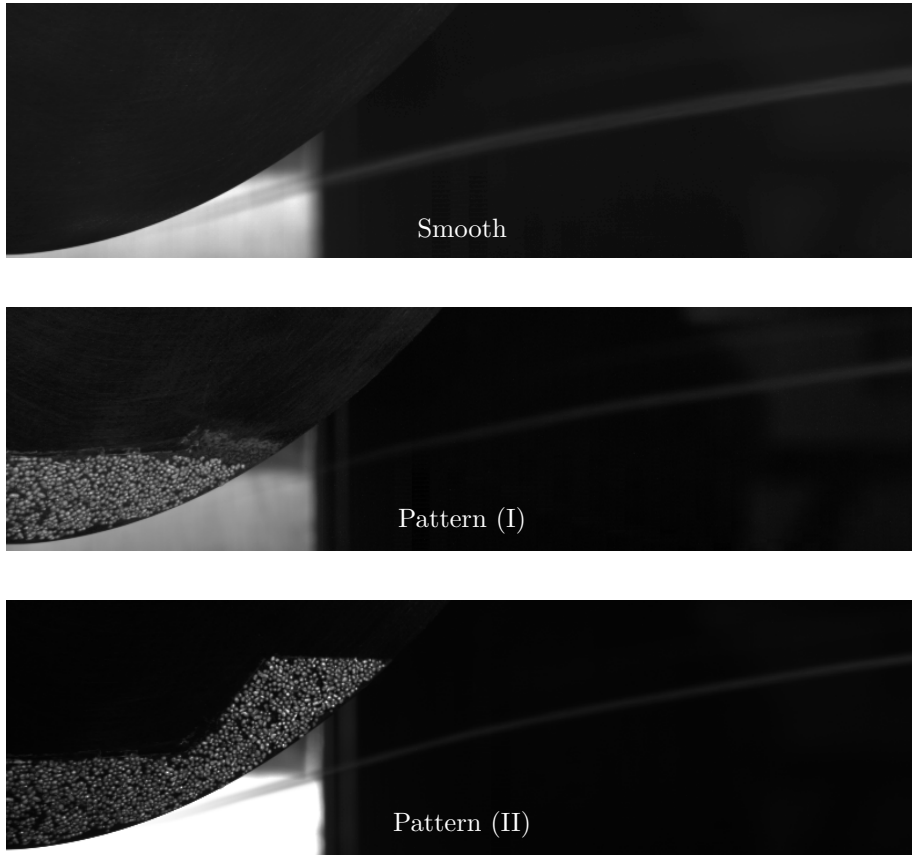

\noindent
The time averaged behavior of the cavitating tip vortex derived from averaging high-speed images over time is presented in Fig. \ref{fig::timeAveCTV}. As it is discussed before, the figure shows that the cavitating tip vortex radius of the smooth foil is larger than of other conditions, indicative of having a stronger tip vortex in this condition. Moreover, for the pattern (I) condition and close to the foil, the cavitating tip vortex is very weak which relates to the discussion provided for Fig. \ref{fig::SSLEroughCavTVTransient}.

\noindent
In Table \ref{table:varLifDrgaCavTVRoughArea}, lift and drag coefficients of different conditions are presented. As expected, the measurements indicate that having roughness will lead to a higher drag. However, the increase depends on the roughness pattern and the way the roughness elements are distributed. For the roughness pattern (II), it is noted that having a non-uniform roughness elements distribution leads to a higher drag compared to the uniform distribution. It should be noted that no obvious difference in the cavitation behaviour or inception property is found for the uniform and non-uniform roughness distribution for pattern (II). This demands for more tests and analyses to clarify the impact of roughness elements in details. The variation of lift is found to be smaller, less than 0.2 $\%$, where in the pattern (I) the lift is increased compared to the smooth foil condition.
\begin{table}[h!]
	\centering  
	\caption{Variation of the drag and lift coefficients and their related standard deviations (SD)for different surface conditions, $\sigma=2.6$, Re=$8.55 \times 10^5$, $\alpha$= 9$^\circ$. }
\begin{tabular}{lccccc}
\hline
              & Smooth  & (I)     & (II)-Non Uniform & (II)-Uniform & (II)-Sparse \\ \hline  \hline
$\text{C}_\text{d}$                 & 0.05420 & 0.05478 & 0.05557          & 0.054540     & 0.05492     \\ 
SD of $\text{C}_\text{d}$ & 0.00329 & 0.00577 & 0.01206          & 0.01054      & 0.01302     \\ 
$\text{C}_\text{l}$                 & 0.62875 & 0.63070 & 0.62837          & 0.62749      & 0.62768     \\ 
SD of $\text{C}_\text{l}$ & 0.00310 & 0.00495 & 0.00871          & 0.00882      & 0.01063     \\ 
Number of samples  & 7       & 4       & 5                & 7            & 6           \\ \hline \hline
\end{tabular}
\label{table:varLifDrgaCavTVRoughArea}
\end{table}

\subsection{Roughness impact on cavitation extent}
\begin{figure} [h!]	
        \centering
        \begin{subfigure}[b]{0.45\textwidth}
        		  \includegraphics[width=\textwidth]{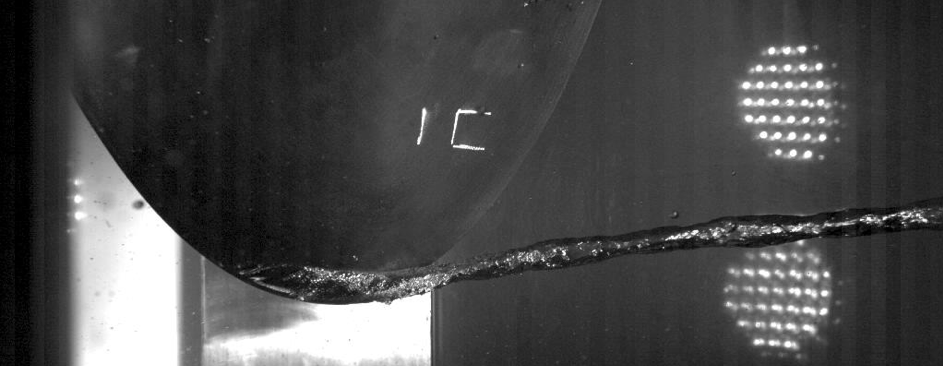}
				  \includegraphics[width=\textwidth]{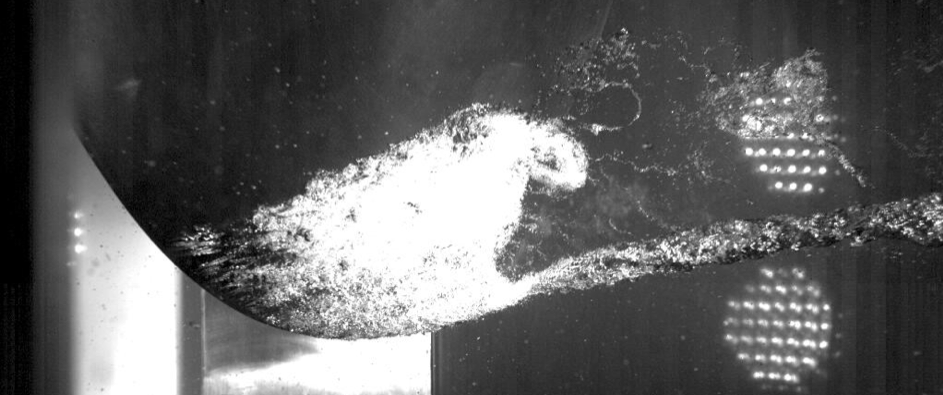}
				  \caption{Smooth}
                  \label{fig::SmoothSigma12MaxMin}
        \end{subfigure}
        \begin{subfigure}[b]{0.45\textwidth}
				  \includegraphics[width=\textwidth]{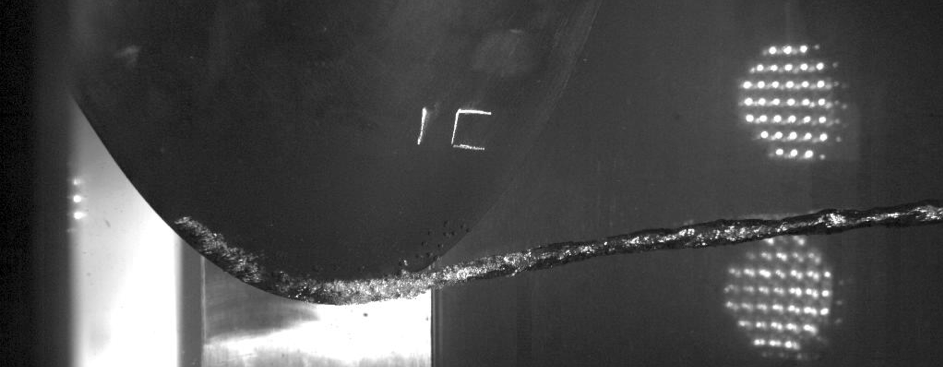}        
				  \includegraphics[width=\textwidth]{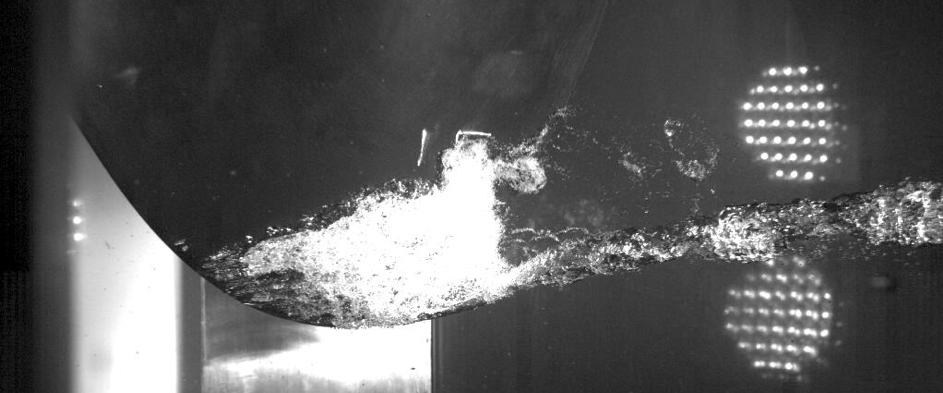}
				  \caption{Pattern (II) - Sparse}
                  \label{fig::tipRoughSigma12MaxMin}
        \end{subfigure}  
		\caption{Minimum and maximum cavitation extent for the smooth and coarse roughness arrangement at $\sigma=1.2$, Re=$1.204 \times 10^6$, $\alpha$= 9$^\circ$.}          
		 \label{fig::SmoothRoughSigma12MaxMin}                   	
\end{figure}

 \noindent
One of the main concerns with roughness application is its impact on the cavitation extent and its regime, especially if roughness triggers the sheet cavity into a more erosive cloudy regime. To investigate this impact, different measurements are conducted in different operating conditions and on different roughness patterns. The discussion here, however, is limited to the comparison of the smooth and roughness pattern (II) having sparse sand distribution. Representative images of cavitation extent of these two patterns are presented in Fig. \ref{fig::SmoothRoughSigma12MaxMin}. At the selected cavitation number, $\sigma=1.2$, the cavitation behaviour shows a relatively constant and strong cavitating tip vortex while on the foil a periodic interaction of sheet and cloud cavitation are observed. The presented figures include the minimum and maximum cavitation extents during this periodic behaviour. 

\noindent
The comparison of the pictures related to the minimum cavitation extent indicates roughness slightly increases the extent of cavitation on the leading edge. This affects the sheet cavity extent on the foil tip and leads to a smaller sheet cavity for the roughened foil. The most interesting effect of roughness on the cavitation behaviour can be noted by the comparison of the maximum cavitation extent. The presence of roughness either prevents the full development of sheet cavity or blocks the re-entrant flow which eventually leads to even less severe cloud cavitation. Both the extent of cloud cavity and its harshness are observed to be reduced on the roughened foil.

\subsection{Roughness and cavitation hysteresis}
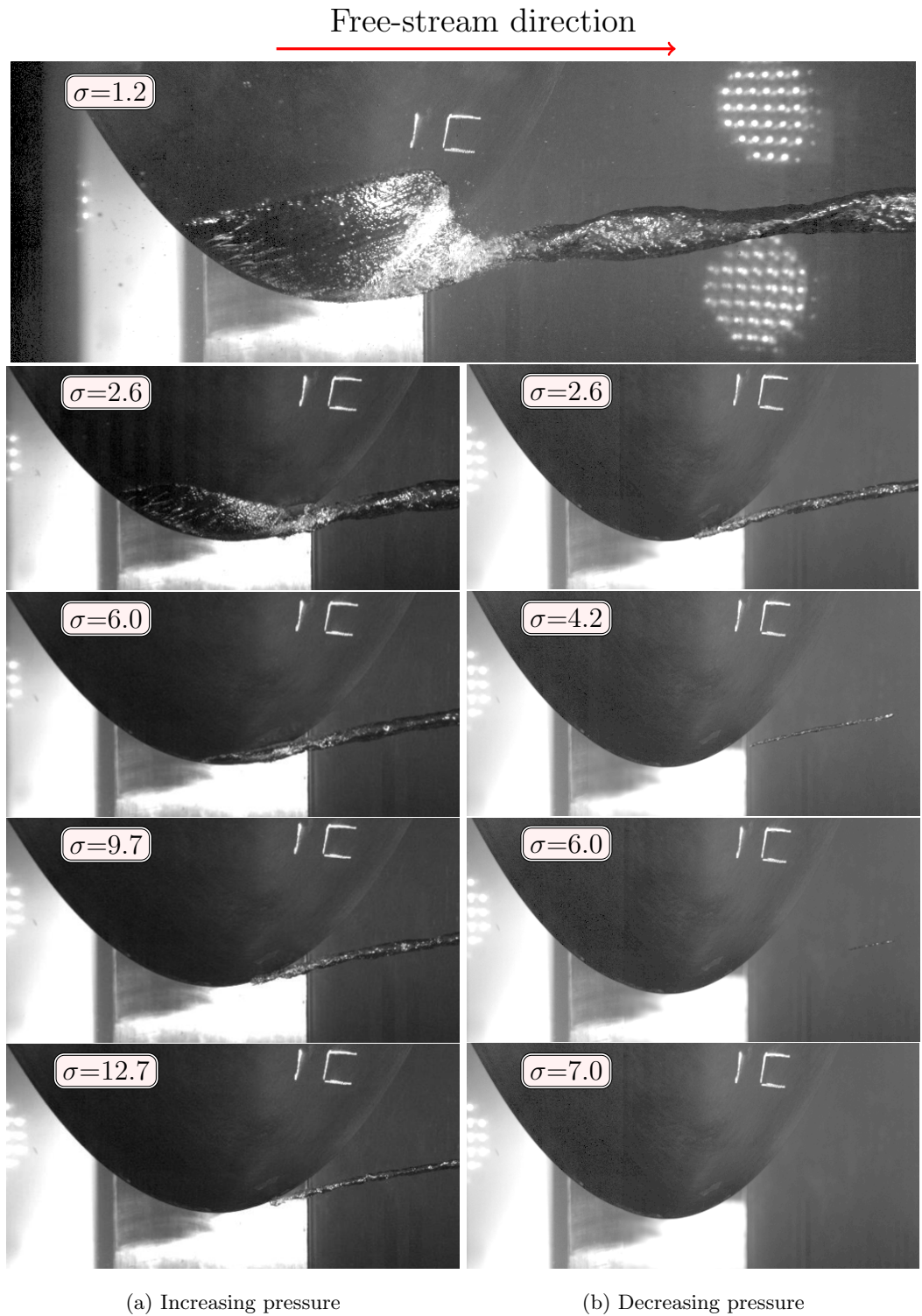
\begin{figure} [h!]	
        \centering
        \begin{subfigure}[b]{0.8\textwidth}
        		  \begin{tikzpicture}
				  \node [anchor=south west,inner sep=0](mesh)at(0,0){
				  \includegraphics[width=\textwidth]{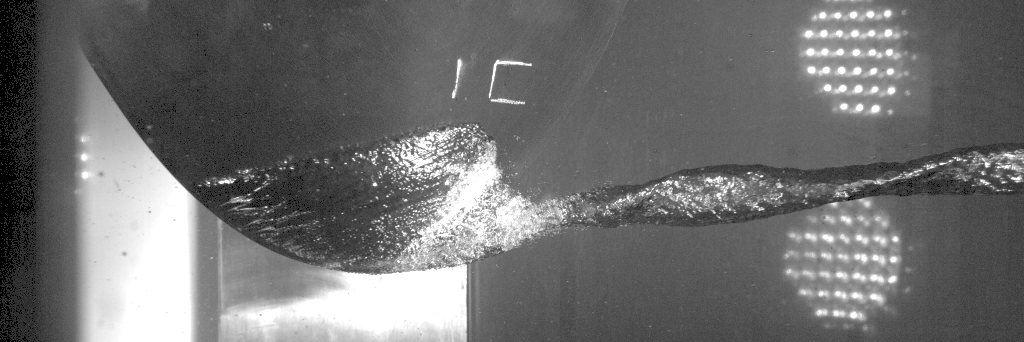} 
									};
					\fill[fill=yellow] (1.5,4.1) node [fill=red!5,draw,double,rounded corners] {\large $\sigma$=1.2};
					
					\node [anchor=south west,inner sep=0, black] (mesh) at (4.8,5.0) {\Large Free-stream direction};	
					\draw [red, very thick, ->]  (4.0,4.75)  ->  (10,4.75);  
				  \end{tikzpicture}	 
        \end{subfigure}            
        \begin{subfigure}[b]{0.4\textwidth}
				  \begin{tikzpicture}
				  \node [anchor=south west,inner sep=0](mesh)at(0,0){
				  \includegraphics[width=\textwidth]{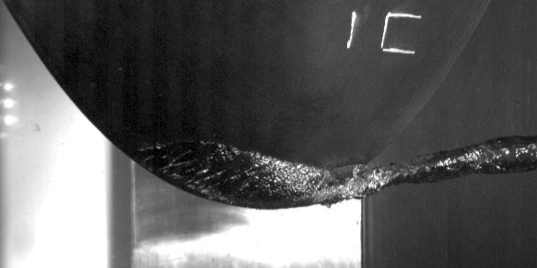} 
									};
					\fill[fill=yellow] (1.5,3.0) node [fill=red!5,draw,double,rounded corners] {\large $\sigma$=2.6};
				  \end{tikzpicture}	         		  
				  \begin{tikzpicture}
				  \node [anchor=south west,inner sep=0](mesh)at(0,0){
				  \includegraphics[width=\textwidth]{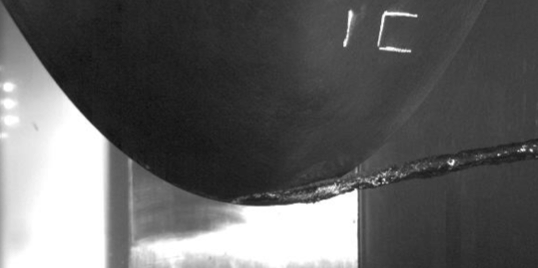} 
									};
					\fill[fill=yellow] (1.5,3.0) node [fill=red!5,draw,double,rounded corners] {\large $\sigma$=6.0};
				  \end{tikzpicture}	 
        		  \begin{tikzpicture}
				  \node [anchor=south west,inner sep=0](mesh)at(0,0){
				  \includegraphics[width=\textwidth]{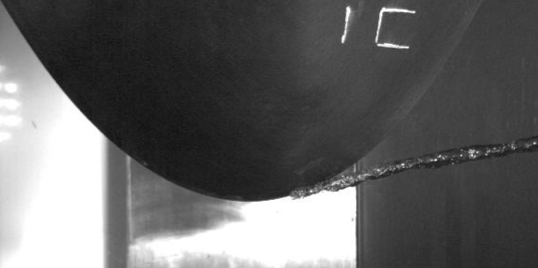} 
									};
					\fill[fill=yellow] (1.5,3.0) node [fill=red!5,draw,double,rounded corners] {\large $\sigma$=9.7};
				  \end{tikzpicture}	 
        		  \begin{tikzpicture}
				  \node [anchor=south west,inner sep=0](mesh)at(0,0){
				  \includegraphics[width=\textwidth]{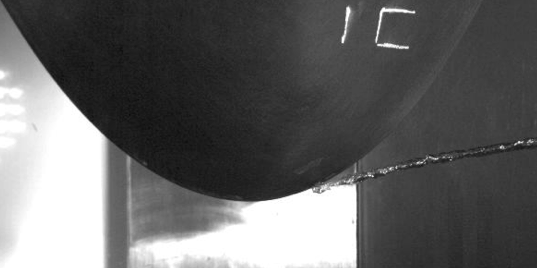} 
									};
					\fill[fill=yellow] (1.5,3.0) node [fill=red!5,draw,double,rounded corners] {\large $\sigma$=12.7};
				  \end{tikzpicture}	 				  
				  \caption{Increasing pressure}
                  \label{fig::SmoothCavHysteresisIncrease}
        \end{subfigure}
        \begin{subfigure}[b]{0.4\textwidth} 
				  \begin{tikzpicture}
				  \node [anchor=south west,inner sep=0](mesh)at(0,0){
				  \includegraphics[width=\textwidth]{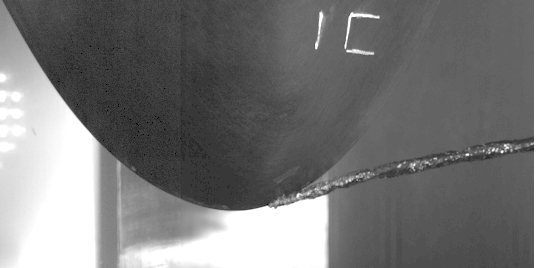} 
									};
					\fill[fill=yellow] (1.5,3.0) node [fill=red!5,draw,double,rounded corners] {\large $\sigma$=2.6};
				  \end{tikzpicture}	         		  
				  \begin{tikzpicture}
				  \node [anchor=south west,inner sep=0](mesh)at(0,0){
				  \includegraphics[width=\textwidth]{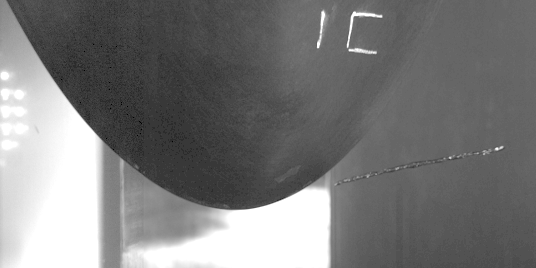} 
									};
					\fill[fill=yellow] (1.5,3.0) node [fill=red!5,draw,double,rounded corners] {\large $\sigma$=4.2};
				  \end{tikzpicture}	 
        		  \begin{tikzpicture}
				  \node [anchor=south west,inner sep=0](mesh)at(0,0){
				  \includegraphics[width=\textwidth]{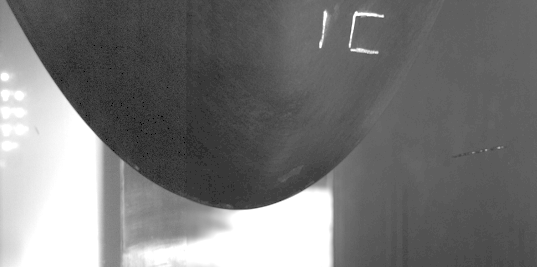} 
									};
					\fill[fill=yellow] (1.5,3.0) node [fill=red!5,draw,double,rounded corners] {\large $\sigma$=6.0};
				  \end{tikzpicture}	 
        		  \begin{tikzpicture}
				  \node [anchor=south west,inner sep=0](mesh)at(0,0){
				  \includegraphics[width=\textwidth]{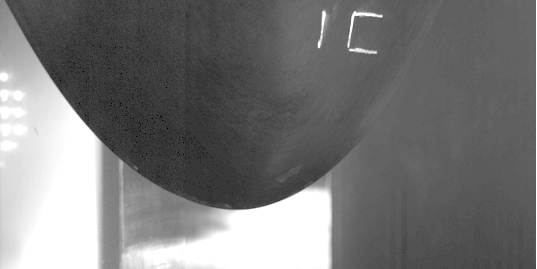} 
									};
					\fill[fill=yellow] (1.5,3.0) node [fill=red!5,draw,double,rounded corners] {\large $\sigma$=7.0};
				  \end{tikzpicture}	 				  
				  \caption{Decreasing pressure}
                  \label{fig::SmoothCavHysteresisDecrease}
        \end{subfigure}        
		\caption{Cavitation hysteresis around the smooth foil, zoomed view of the suction side tip.}          
		 \label{fig::SmoothCavHysteresis}                   	
\end{figure}

\begin{figure} [h!]	
        \centering
        \begin{subfigure}[b]{0.788\textwidth}
        		  \begin{tikzpicture}
				  \node [anchor=south west,inner sep=0](mesh)at(0,0){
				  \includegraphics[width=\textwidth]{Figure/CavitatingTV/sigma1.2/1a.png}				  
									};
					\fill[fill=yellow] (1.5,4.1) node [fill=red!5,draw,double,rounded corners] {\large $\sigma$=1.2};
										
				  \end{tikzpicture}	 
        \end{subfigure}
                
        \begin{subfigure}[b]{0.38\textwidth}
				  \begin{tikzpicture}
				  \node [anchor=south west,inner sep=0](mesh)at(0,0){
				  \includegraphics[width=\textwidth]{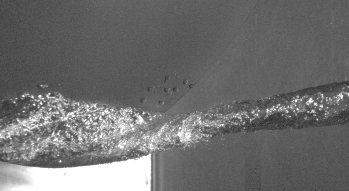} 
									};
					\fill[fill=yellow] (1.5,3.0) node [fill=red!5,draw,double,rounded corners] {\large $\sigma$=2.6};
				  \end{tikzpicture}	         		  				  
				  \caption{Increasing pressure}
                  \label{fig::RoughCavHysteresisIncrease}
        \end{subfigure}
        \quad
        \begin{subfigure}[b]{0.38\textwidth} 
				  \begin{tikzpicture}
				  \node [anchor=south west,inner sep=0](mesh)at(0,0){
				  \includegraphics[width=\textwidth]{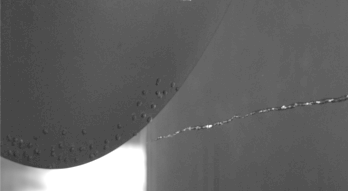} 
									};
					\fill[fill=yellow] (1.5,3.0) node [fill=red!5,draw,double,rounded corners] {\large $\sigma$=2.6};
				  \end{tikzpicture}	         		  
				  \caption{Decreasing pressure}
                  \label{fig::RoughCavHysteresisDecrease}
        \end{subfigure}        
		\caption{Cavitation hysteresis around the roughened foil, zoomed view of the suction side tip.}          
		 \label{fig::RoughCavHysteresis}                   	
\end{figure}

 \noindent
In Fig. \ref{fig::SmoothCavHysteresis} and Fig. \ref{fig::RoughCavHysteresis}, the hysteresis of cavitation are investigated on the smooth foil and for roughness pattern (II) with sparse sand distribution. The figures contain two columns related to the conditions where the cavitation tunnel pressure is gradually decreased or increased while keeping the tunnel free-stream velocity constant. 

\noindent
For the smooth foil, a significant difference is found between the cavitation behaviour of the decreasing and increasing tunnel pressure. While decreasing the pressure, cavitation incepts at around $\sigma$=6 and further decrease leads to a stronger TVC. At $\sigma$=1.2, the cavitation contains both a TVC and sheet/cloud cavitation. By increasing the tunnel pressure, the sheet/cloud cavitation disappears but the TVC remains attached to the tip until $\sigma$=12.7. At this pressure condition, TVC suddenly separates from the tip and after that no further TVC is observed. For the roughened foil, the hysteresis analysis only includes $\sigma$=1.2 and $\sigma$=2.6. This, however, was enough to indicate that the roughness does not worsen the cavitation hysteresis in the variable pressure condition compared to the smooth foil. 

\section{Conclusion} 
 This paper demonstrates the possibility of using roughness in order to mitigate cavitating tip vortex flows. Knowing that its tip vortex flow resembles propeller tip vortex flows, an elliptical foil is selected as the test case. The tip vortex properties of this foil at two roughness pattern configurations are evaluated at different operating conditions. The main findings of the study are summarized as follow, 
\begin{itemize}
  \item Application of roughness is effective in delaying vortex cavitation inception. The measurements show a decrease in the TVCI as large as 33 \% in the optimized roughness pattern compared to the smooth foil condition where the drag force increase is observed to be around 2 \%.
  \item Application of roughness on the leading edge, tip region and trailing edge of the suction side is found to be the most optimum configuration of roughness pattern where the TVCI mitigation is achieved with little performance degradation.
    \item We did not observe any risk of increasing sheet cavity on the foil by roughness application in the tested operating conditions. It is, however, noted that roughness elements form tiny nucleation sites having micro- or even nano-metric residual air pockets  that generate very small nuclei continuously and normally due to local degassing.
  \item Roughness application not only changes the tip vortex strength and the inception point but also the free-stream nuclei capture properties of the tip vortex. This can be related to the fact that nuclei movements are affected by more interactive structures generated by the roughness elements.
  \item The TVCI is found to be similar in the uniform and non-uniform roughness distributions for the same roughness pattern while the drag force is found to be higher in the non-uniform roughness distribution. This, however, poses a further question on how roughness elements should be distributed in terms of randomness and populations to minimize the performance degradation. 
  \item The measurements highlight that in the roughened surface condition magnitudes of the tangential and axial velocities are decreased compared to the smooth condition. This clarifies that lower momentum has been fed into the tip vortex in the roughened case. One possible reason for this could be increased velocity fluctuations in the roughened cases compared to the smooth condition which result in higher mixing rates and therefore higher viscous losses.  
  \item For a fully cavitating condition where the foil experiences a periodic combination of cavitating tip vortex and sheet/cloud cavity, observations indicate that roughness either prevents the full development of sheet cavity or blocks the re-entrant flow which eventually leads to less severe cloud cavitation and possibly less erosive cavitation regime in the roughened foil compared to the smooth foil condition.
        
\end{itemize}
These conclusions are obviously to be confirmed for other ranges of the variables that were not tested in this study.

\section{Declarations} 		
 \noindent
\textbf{Funding:} Financial support for this work has been provided by VINNOVA through the RoughProp project, Grant number 2018-04085. The experimental measurements are conducted in the free surface cavitation tunnel at the Kongsberg Hydrodynamic Research Center, Kristinehamn, Sweden.

\noindent
\textbf{Availability of data and material:} Raw data were generated at the cavitation tunnel facility of Kongsberg Maritime AB, Kristinehamn, Sweden. Derived data supporting the findings of this study are available from the corresponding author upon reasonable request.

\noindent
\textbf{Conflicts of interest and Code availability:} Not applicable.  
\newpage	  
\section{Nomenclature} 		
 \begin{tabular}{ l   p{7cm} }
	$\text{C}_0=301.2$ mm   	& root chord length	  					\\ 
	$\text{S}=360$ mm   			& span length	\\
	A=$8.43 \times 10^{-2}$ $\text{m}^2$ & projected surface area of the foil      \\ 
	$\rho$  		        & density   \\
	$\mu$   		        	& dynamic viscosity\\
	$ \text{U}_\text{inlet}$  	& inlet velocity	\\      	
	p$_{\text{outlet}}$  			& tunnel outlet pressure  	\\ 
	p$_{\text{sat}}$  				& saturation pressure\\    
	$\sigma=(\text{p}_{\text{outlet}}-\text{p}_{\text{sat}})/(0.5 \rho \text{U}_{\text{inlet}})$  	& cavitation number \\
	Re=$\rho \text{U}_{\text{inlet}} \text{C}_0 / \mu$   									& Reynolds number \\     
	D  							& drag force \\
	L  							& lift force    \\
	$\text{C}_\text{d}=\text{D}/(0.5 \rho {\text{U}_\text{inlet}}^2 A)$ & drag coefficient \\
	$\text{C}_\text{l}=\text{L}/(0.5 \rho {\text{U}_\text{inlet}}^2 A)$ & lift Coefficient\\
	r  					& radial distance \\
	r$_\text{vortex}$  	& vortex core radius\\
	z  					& streamwise direction, Fig. \ref{fig::coordinateSystem} \\  
	Do	  & dissolved oxygen concentration  \\
	AOA	  & angle of attack \\
	TVC   & tip vortex cavitation \\
	TLP	  & tip loaded propeller \\
	SSLE  & suction side leading edge, Fig. \ref{fig::testedConditionsRoughnessPattern}    \\
	SSTE  & suction side trailing edge, Fig. \ref{fig::testedConditionsRoughnessPattern}   \\	 
	NU  & non-uniform roughness distribution, Fig. \ref{fig::uniformAndNonUniformDisSandGrains}  \\
\end{tabular}

 %   \bibliographystyle{acm}
 %   \bibliography{main}      
     
\end{document}